# Development and Field Validation of a Fully Customised Vehicle Scanning System on Two Full-Scale Bridges

A. Calderon Hurtado[a], J. Xu[a], R. Salleh[b], D. Dias-da-Costa[c], M. Makki Alamdari[a,*]

[a]*Center for Infrastructure Engineering and Safety, School of Civil and Environmental Engineering, University of New South Wales, Sydney, NSW 2052, Australia*
[b]*School of Computer Science and Engineering, University of New South Wales, Sydney, NSW 2052, Australia*
[c]*School of Civil Engineering, The University of Sydney, NSW 2006, Australia*

---

## Abstract

Ensuring the structural integrity of bridges is essential for maintaining infrastructure safety and promoting long-term sustainability. In this context, Indirect Structural Health Monitoring (ISHM) through drive-by bridge inspection emerges as a promising alternative to traditional inspection methods, offering a cost-effective and scalable solution by using vehicle-mounted sensors to assess the condition of bridges without requiring direct instrumentation. This study introduces the first purpose-built electric inspection vehicle specifically designed for drive-by bridge inspection. The autonomous platform is capable of maintaining a constant low speed and offers customisable operational parameters to maximise the accuracy and repeatability of indirect sensing — capabilities not achieved in previous studies. The vehicle is deployed within an ISHM framework and tested on two full-scale bridges to evaluate its effectiveness in capturing structural dynamic responses. Two unsupervised frameworks are then employed to analyse the collected data to identify features indicative of bridge properties and structural condition. The promising findings from this study demonstrate the practical feasibility of the approach. The study also shows the potential of ISHM as a viable tool for efficient bridge monitoring, contributing to the development of next-generation structural health monitoring systems that can enhance safety, optimise maintenance strategies, and support the longevity of critical infrastructure.

---

## 1. Introduction

Bridges are critical components of modern infrastructure, enabling the transportation of people and goods while supporting economic stability and connectivity [1]. Their structural integrity is vital to public safety, operational efficiency, and sustainable development. However, as these structures age, they are subjected to various deteriorating factors, including environmental effects, material degradation, increasing traffic loads, and unexpected external events [2]. Such degradation can lead to catastrophic failures, as seen in the collapse of the Morandi Bridge in Italy (2018) and the failure of the Saint Anthony Falls Bridge in the United States (2007).

*Corresponding author
*Email address:* m.makkialamdari@unsw.edu.au (M. Makki Alamdari)

Worldwide, bridge infrastructure maintenance is an increasing challenge. For instance, in Australia, the majority of bridges were constructed prior to 1976, and many are now approaching or even exceeding their design lifespans [3]. At the same time, bridge usage is projected to increase by 300% over the next three decades, intensifying the demand for reliable and cost-effective monitoring strategies [4]. Currently, periodic visual inspection remains the main preemptive approach, typically conducted every two years [5]. Despite its widespread use, there are several limitations in what concerns visual inspections, such as being subjective, labour-intensive, expensive, and inherently restricted in scope due to its dependence on the inspector [6]. Consequently, the demand for automated, objective, and continuous Structural Health Monitoring Systems (SHMS) has grown considerably [7, 8].

To address these limitations, Indirect Structural Health Monitoring (ISHM) using drive-by bridge inspection has emerged as a promising alternative to visual inspections and conventional SHMS, which rely on fixed sensors installed directly on the structure [9]. In contrast, drive-by bridge inspection involves instrumented vehicles crossing the bridge, using the dynamic response measured by mounted sensors, to infer the condition of the bridge [10]. This approach enables the extraction of modal properties, such as natural frequencies, damping characteristics, and potential damage indicators. Its advantages include low cost, particularly in the face of scalability across many structures, and reduced deployment and maintenance effort [2].

Over the past two decades, numerous laboratory studies have contributed to validating drive-by methodologies under controlled conditions. Zhang et al. [11] used wavelet-based time-frequency analysis to extract mode shapes from a scaled beam, while Kim et al. [12] introduced artificial damage through saw cuts and added masses to demonstrate damage identification using frequency and damping changes. Cerda et al. [5, 13] conducted scaled laboratory tests where damage was simulated by changing boundary conditions and adding mass, validating classification-based identification techniques using frequency-domain analysis. Mei et al. [14] applied Mel-frequency spectral coefficients (MFCC) and Principal Component Analysis (PCA) to detect damage in a scaled setup using a two-axle instrumented vehicle.

Recent studies have focused on improving robustness under variable operating conditions. Liu and Xu [15] proposed a multi-task unsupervised learning framework that uses Short-Time Fourier Transform (STFT) signals from multiple vehicles to identify, localise, and quantify bridge damage, achieving high identification accuracy across different test scenarios. Li et al. [16] combined MFCCs with Support Vector Machines (SVM) to achieve high detection sensitivity and low computational cost in a scaled experiment. Singh and Sadhu [17] introduced a hybrid signal processing technique that effectively isolates bridge dynamics from noisy vehicle signals using Wavelet Packet Transform (WPT) and Synchro-Extracting Transform (SET), even under varying speeds. Yang et al. [18] demonstrated the effectiveness of Variational Mode Decomposition (VMD) in extracting modal frequencies despite strong vehicle-induced noise.

In terms of model generalisation and scalability, Liu et al. [19] developed a transfer learning method using adversarial and multi-task networks, allowing damage classifiers to generalise across different structures without labelled data. Matarazzo et al. [20] introduced STRIDEX, a blind deconvolution approach that isolates bridge responses from vehicle signals without relying on detailed system models. Sadeghi et al. [21] proposed the Crowdsourced Modal Identification using Continuous Wavelets (CMICW) technique, showing that fusing data from multiple smartphone-equipped vehicles increases identification reliability compared to single-vehicle systems.

Although laboratory validations are promising, real-world applications are more complex and remain limited. Most experimental tests in real bridges focus on the identification of dynamic properties. Yang et al. [22] and Lin et al. [23] extracted bridge natural frequencies through frequency-based methods, with the latter using additional truck mass to enhance response amplitude. Locke et al. [24] applied Operational Modal Analysis (OMA) in a field setting, though their methodology was constrained to short-span bridges. Yang et al. [25] conducted field testing on the Turtle Mountain Bridge using frequency-domain analysis of contact point accelerations to extract dynamic properties, providing practical validation of vehicle-bridge interaction models. Additional field validations by Kong et al. [26] and Xu et al. [27] confirmed the effectiveness of tractor-trailer and contact-point acceleration approaches, respectively, for identifying bridge frequencies without fixed sensors. McGetrick et al. [28] emphasised the challenges of simulating repeatable artificial damage in operational bridges, which remains a key limitation for full-scale ISHM validation.

Despite these successes, ongoing research in advanced signal processing, robust machine learning techniques, and optimised vehicle configurations continues to enhance the feasibility of these methods for real-world applications.

This study advances drive-by bridge inspection for full-scale, operational structures by building on two rigorous ISHM frameworks [29, 30, 31]. For the first time, we introduce a purpose-built autonomous inspection vehicle and demonstrate its application in field tests to identify key dynamic properties and evaluate its practical potential for damage assessment. The outcomes of this work contribute to the development of more scalable, cost-effective structural health monitoring systems that support the long-term safety, maintenance, and sustainability of bridge infrastructure.

## 2. Research Significance

Building on previous work by this research group [29, 30, 31], which introduced novel drive-by bridge damage assessment frameworks validated under controlled laboratory conditions, our study extends those unsupervised approaches to two full-scale, in-service bridges. The primary contributions of this research are as follows:

a) *Development of an Autonomous Inspection Vehicle*: We designed, constructed, and field-characterised the first purpose-built autonomous platform for drive-by bridge inspection. This platform integrates custom instrumentation and control systems to serve as a versatile mobile sensing unit.

b) *Indirect Characterisation of Bridge Dynamics*: During multiple crossings over real bridges, the autonomous vehicle successfully captured vibrational data that revealed key dynamic properties (e.g., natural frequencies) without the need for fixed sensors on the structures.

c) *Field Validation of Unsupervised Damage Assessment Techniques*: The study applied and validated previously proposed indirect sensing methodologies under operational conditions, demonstrating their effectiveness for real-world damage detection and highlights the practical value of prior ISHM frameworks [29, 30, 31].

The paper is structured as follows: Section 3 details the configuration and characterisation of the proposed novel inspection vehicle. Section 4 presents the two case studies used for field test validation. Sections 5 and 6, respectively, describe the proposed methodologies for damage assessment, as well as the results obtained from both direct and indirect sensing across the two case studies. Finally, Section 7 summarises the main findings and provides recommendations for future work.

## 3. A Novel Autonomous Inspection Vehicle

This section presents the inspection vehicle setup, specifically designed and developed to perform drive-by bridge inspections. The inspection vehicle is a customised 1:5 scale electric RC car, adapted to carry the necessary instrumentation for bridge monitoring. Figure 1 illustrates the complete inspection vehicle and its data management system. Specifically, it is equipped with four PCB-ICP393B05 accelerometers, one mounted on the suspension system of each wheel, capable of measuring within a ±0.5 g range and operating across a frequency range of 0.7 to 450 Hz. These sensors effectively capture vehicle dynamic responses, which are crucial for assessing bridge conditions. The setup also includes four Honeywell Model 31 Series load cells (capacity 0.5–5,000 N), mounted on a custom-designed aluminium suspension tower. This frame not only houses the sensors but also allows adjustment of the pivoting point to accommodate various load configurations (see Figure 2b). The sensing data are collected using a NI cDAQ-9185 CompactDAQ system and transmitted wirelessly to a remote computer via a 300M PIX-LINK WR09Q wireless repeater. The data are accessed through NI MAX and recorded using NI-FlexLogger, ensuring a reliable workflow for acquisition, processing, and storage of dynamic measurements.

The autonomous control architecture of the vehicle ensures both precise path following and a stable inspection speed, critical for consistent data collection for drive-by bridge scanning purposes. Steering guidance is achieved using a RoboteQ MGSW1600 magnetic guide sensor, which continuously tracks a magnetic tape embedded along the inspection route. Its real-time output interfaces directly with the steering servo, enabling precise path tracking and delivering high accuracy in lateral positioning, while also facilitating smooth navigation even over uneven surfaces. A closed-loop feedback system handles speed regulation: an AMS Hall-effect sensor mounted on the drive train shaft measures rotational speed and relays this data to an ESP32 microcontroller. The ESP32 implements a PID control algorithm that adjusts the motor speed via a standard R/C electronic speed controller (ESC), working in tandem with a reduction gearbox to convert high-RPM motor output into the slow, torque-rich speeds required for scanning. A double-pole double-throw (DPDT) switch provides seamless toggling between manual and autonomous modes for both steering and motor drive control, allowing the operator to intervene for greater flexibility in operation, if required. This integrated system maintains target speeds, maximises data repeatability, and minimises operator workload during extended field campaigns. This architecture is shown in

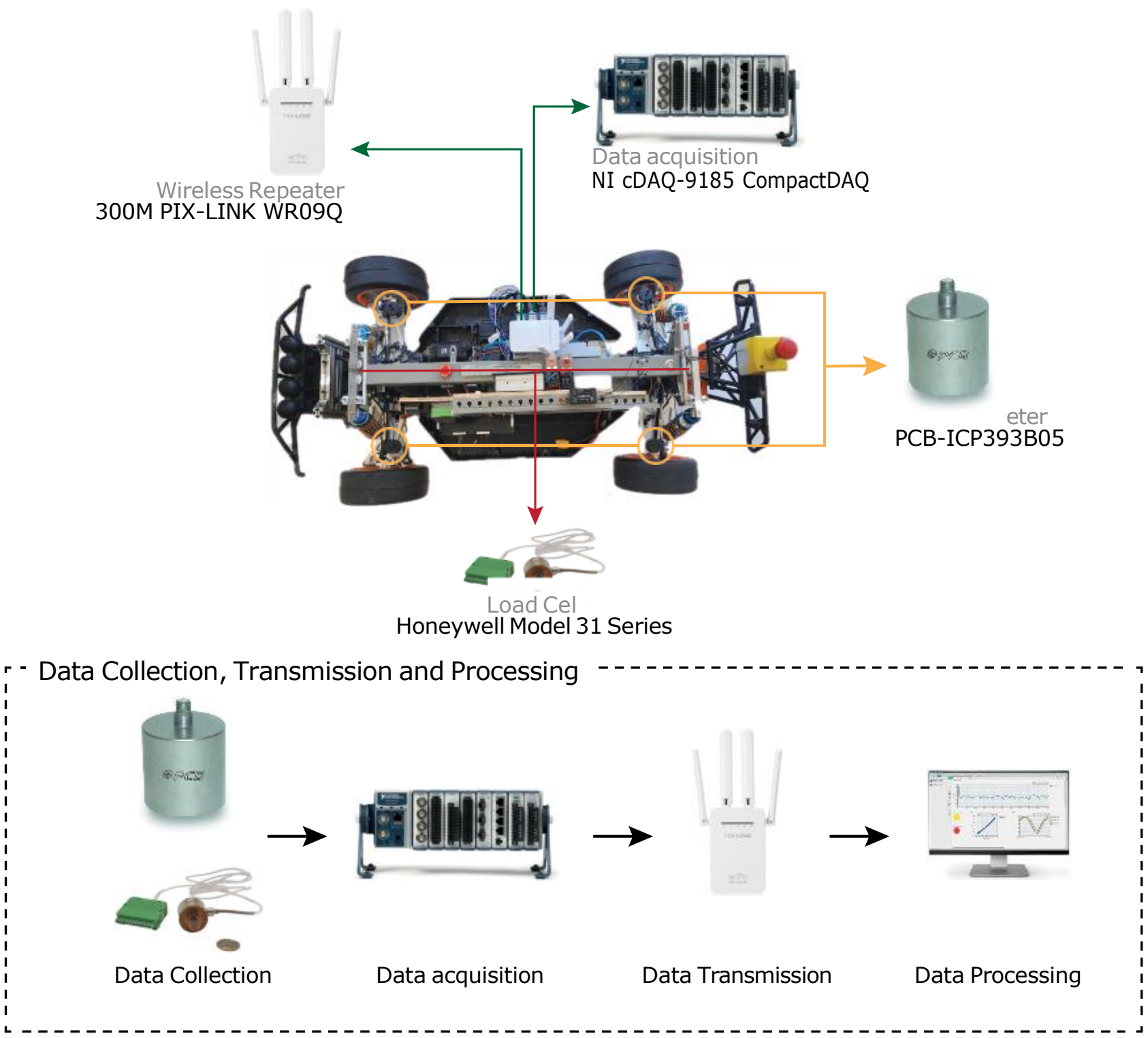

Figure 1: Inspection vehicle setup for drive-by bridge inspection.

Figure 2a. The overall system, with a vehicle mass of 20 kg and axle distance of 80 cm, offers a robust and adaptable platform for efficient and repeatable drive-by bridge inspections, with operating speeds varying from 0.1 m/s to 10 m/s.

### 3.1. *Characteristics of the Inspection Vehicle*

Accurately characterising the dynamic properties of the inspection vehicle is essential for reliable drive-by sensing. To this end, a driving test is performed in which the vehicle is driven at a constant velocity of 0.17 m/s for 20 seconds to estimate the frequencies associated with the electric motor under operational conditions. Acceleration responses from four accelerometers mounted on the suspension are recorded and processed using Frequency Domain Decomposition (FDD) to identify the dominant motor-related frequencies.

The FFD methodology is expressed as follows [32]: consider a matrix $A \in \mathrm{IR}^{n\times m}$, where $n$ is the number of observations in the acceleration signals and $m$ is the number of sensors, the cross-power spectral density (CPSD), $\mathbf{S}$, computed at each frequency is given as:

$$\mathbf{S}(f) = \begin{bmatrix} S_{11}(f) & S_{12}(f) & \cdots & S_{1m}(f) \\ S_{21}(f) & S_{22}(f) & \cdots & S_{2m}(f) \\ \vdots & \vdots & \ddots & \vdots \\ S_{m1}(f) & S_{m2}(f) & \cdots & S_{mm}(f) \end{bmatrix} \in \mathbb{C}^{m\times m}, \tag{1}$$

where $f$ is the vector of discrete frequencies, and $S_{ij}(f)$ is the complex-valued cross-power spectral density between signals $i$ and $j$, with $i, j = 1, 2, ..., m$. Once the CPSD is obtained,

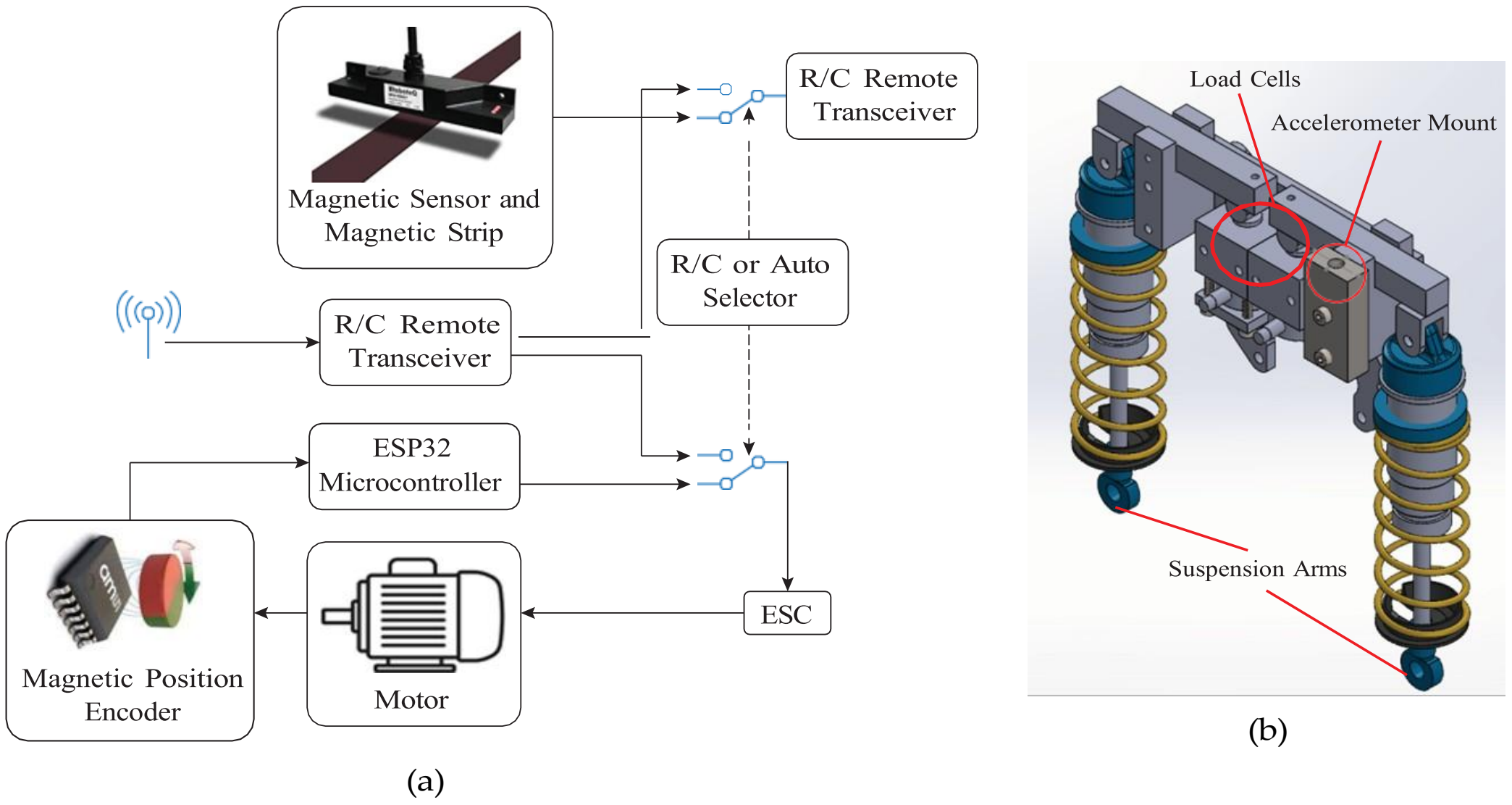

Figure 2: Illustration of: (a) system block diagram; and (b) suspension tower.

Singular Value Decomposition (SVD) can be derived for each frequency in the spectrum as in the following equation:

$$\mathbf{S} = \mathbf{U}\,\boldsymbol{\Sigma}\,\mathbf{U}^{H}, \tag{2}$$

where $\mathbf{U}$ contains the mode-shapes, $\boldsymbol{\Sigma}$ is a diagonal matrix containing the singular values, and $H$ denotes the Hermitian transpose. The dominant peaks observed in the SVD diagram correspond to the frequencies of the car in operation.

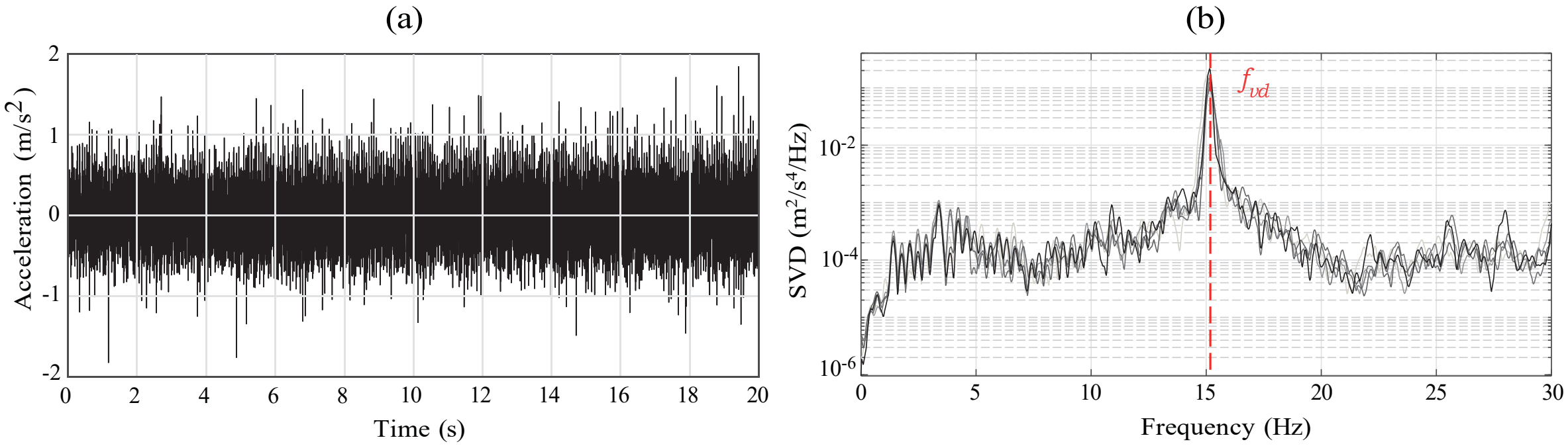

Figure 3: Identification of the driving frequency associated with the electric motor in the inspection vehicle: (a) acceleration time response from the accelerometer mounted on the front-right suspension arm; (b) the first singular value from five different driving tests incorporating all four accelerometer responses. The red dashed line represents $f_{vd}$.

Figure 3 presents the analysis of the test conducted to characterise the inspection vehicle. Figure 3a shows an example of the acceleration response from one of the sensors during a driving test. Figure 3b illustrates the first singular value, from the FDD analysis, across five different tests. It can be observed that the electric motor produces a dominant frequency of 15 Hz. This frequency is identified in this work as $f_{vd}$. This information is crucial for analysing the recorded data and identifying the dynamic properties of the bridge. It should be emphasised that $f_{vd}$ depends on several factors, including the speed of the vehicle, battery level and power; however, according to our investigations, it will vary between 15 $\pm$2 Hz.

# 4. Experimental Case Studies

## *4.1. Case Study I: Pedestrian Bridge at UNSW*

The first case study analyses a pedestrian bridge at the Kensington campus of the University of New South Wales in Sydney, Australia. The bridge is a simply supported steel structure with a total length of 17 m and a width of 80 cm (see Figure 4).

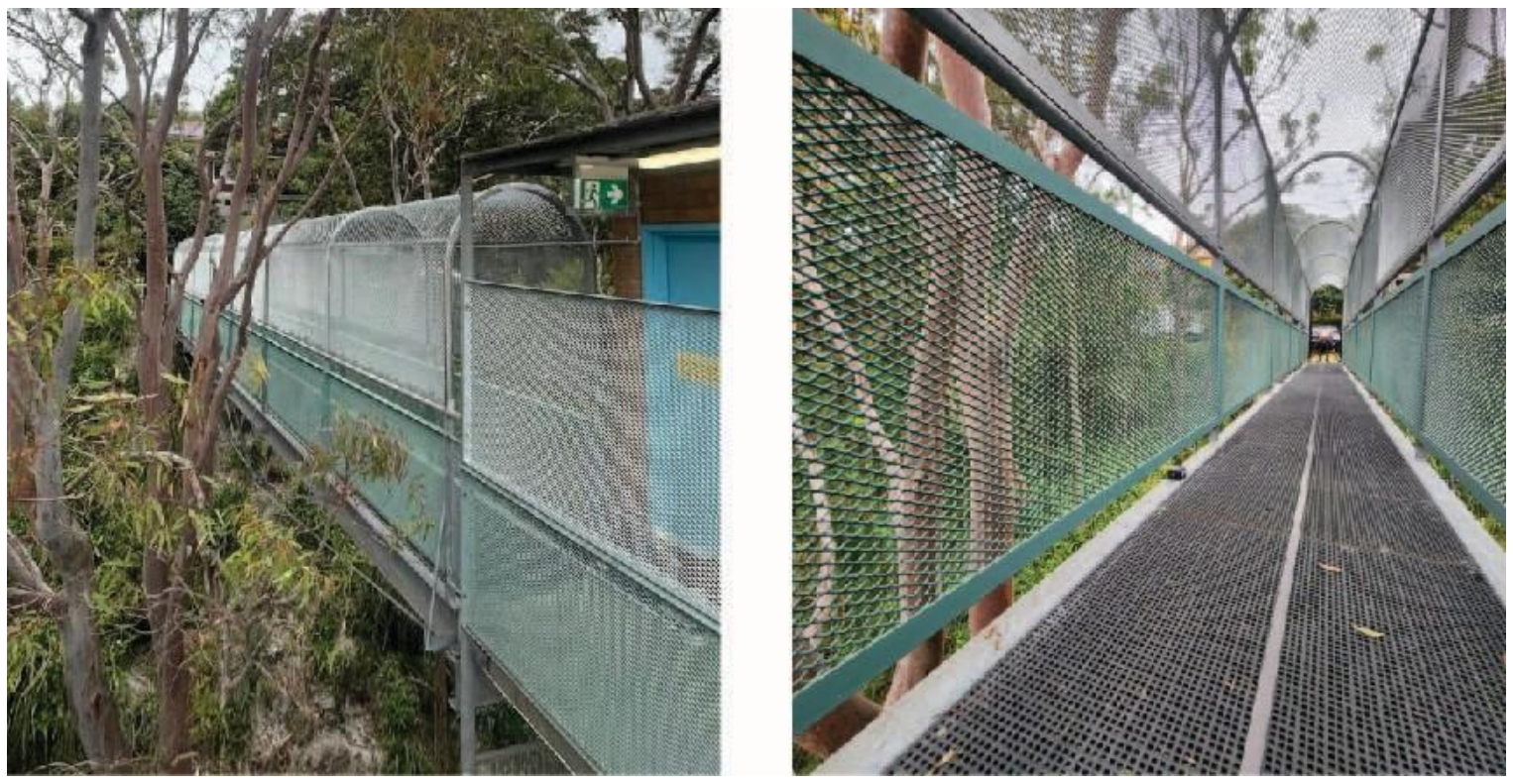

Figure 4: Pedestrian bridge to be inspected, located at the Kensington campus at the University of New South Wales, Australia.

Two measurement campaigns are conducted on this pedestrian bridge – see Figure 5. First, a direct sensing is conducted to identify the first natural frequency of the structure, which is later used as a benchmark to verify the indirect sensing. In the second stage, an indirect sensing using drive-by inspection is performed on the bridge.

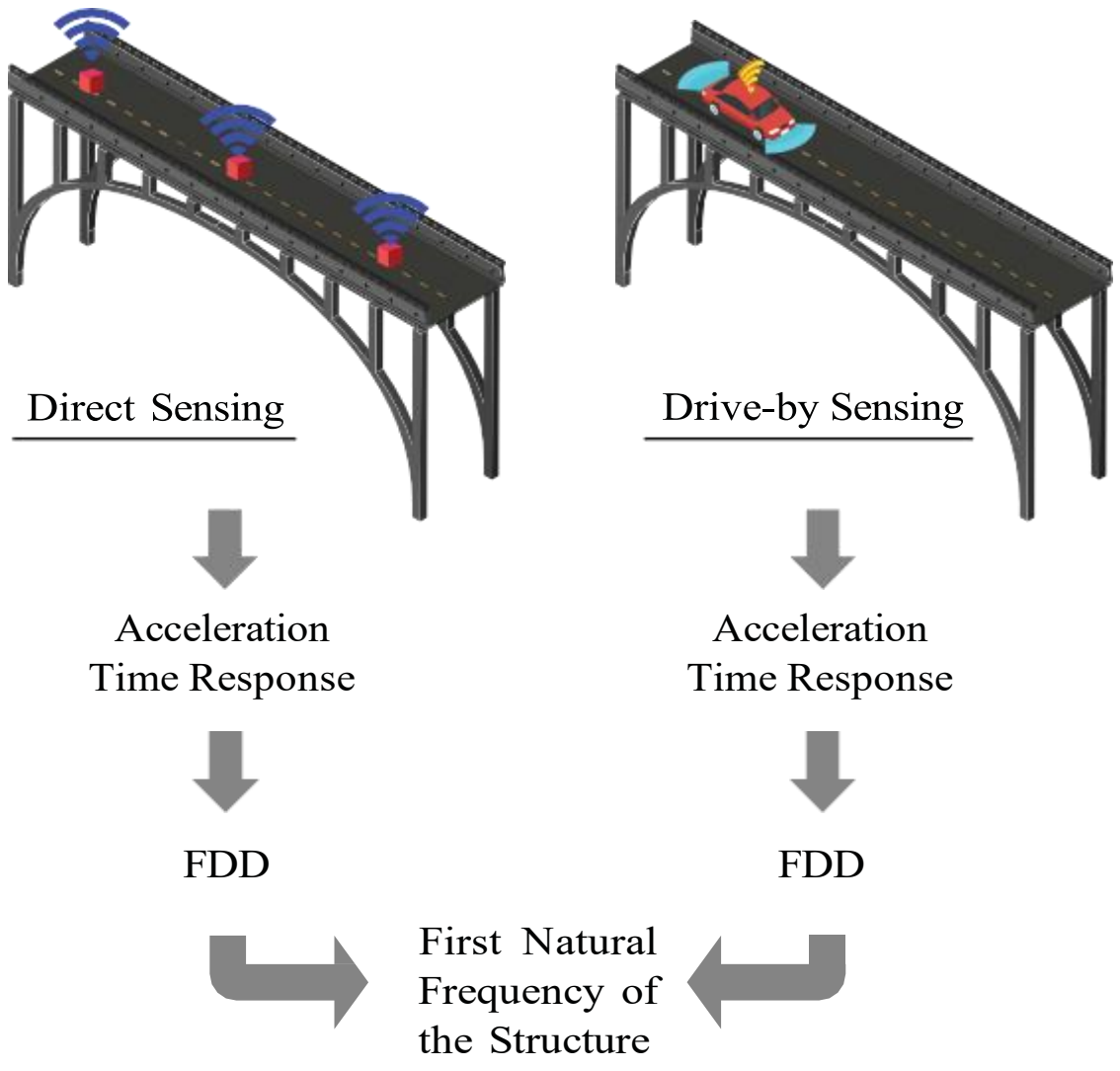

Figure 5: Natural frequency identification of the bridge.

#### 4.1.1. Direct Sensing

For the first test, three wireless accelerometers are installed at the quarter-span locations of the bridge, as shown in Figure 6a. The sensors used are BeanAir Wilow AX-3D triaxial accelerometers, configured with a sampling frequency of 500 Hz. During the test, 30 samples of 15 seconds each are collected while the structure is subjected to an external excitation induced by pedestrians.

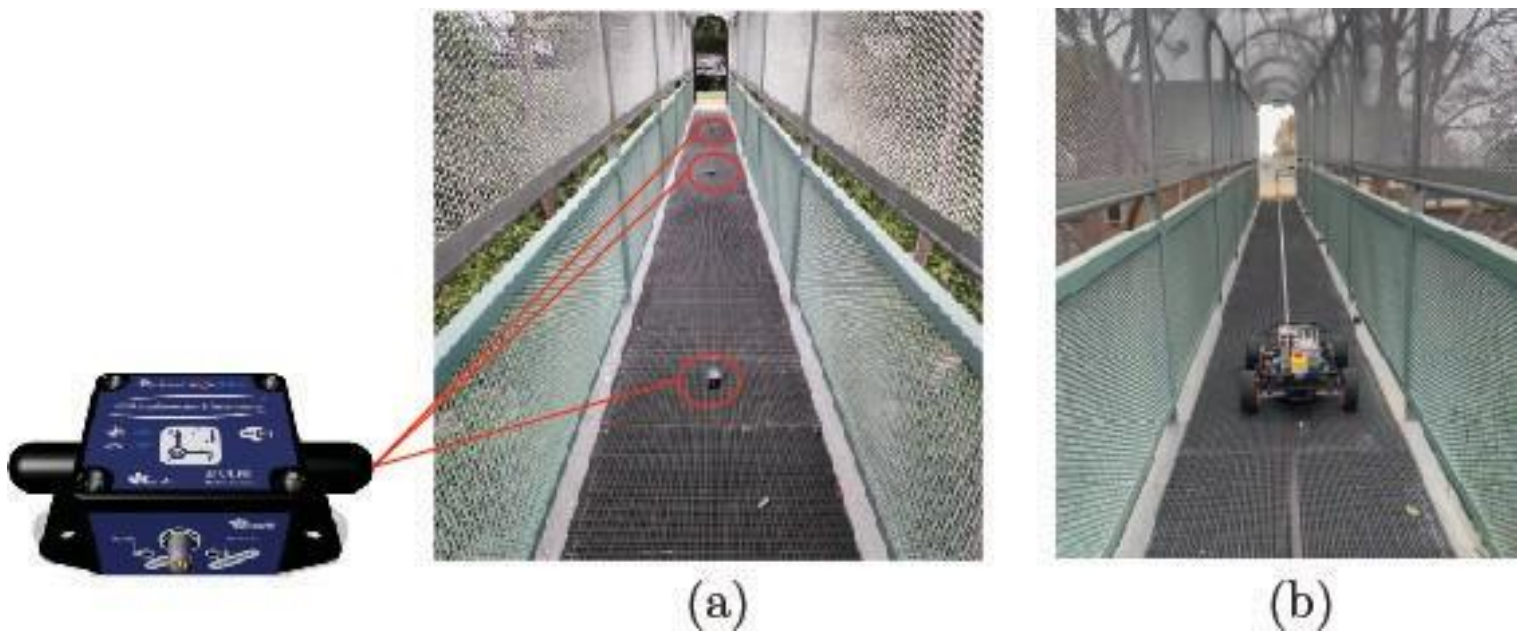

Figure 6: Tests on the Pedestrian Bridge at UNSW: (a) direct test with three triaxial BeanAir sensors directly installed on the bridge; and (b) drive-by inspection.

To identify the first natural frequency, the FDD method is applied to the acceleration signals. As described in Section 3.1, the FDD method involves computing the SVD of the CPSD matrix constructed from the acceleration signals. In this study, only the first singular value is considered, and the peak in this spectrum is taken as the estimate of the first natural frequency of the structure.

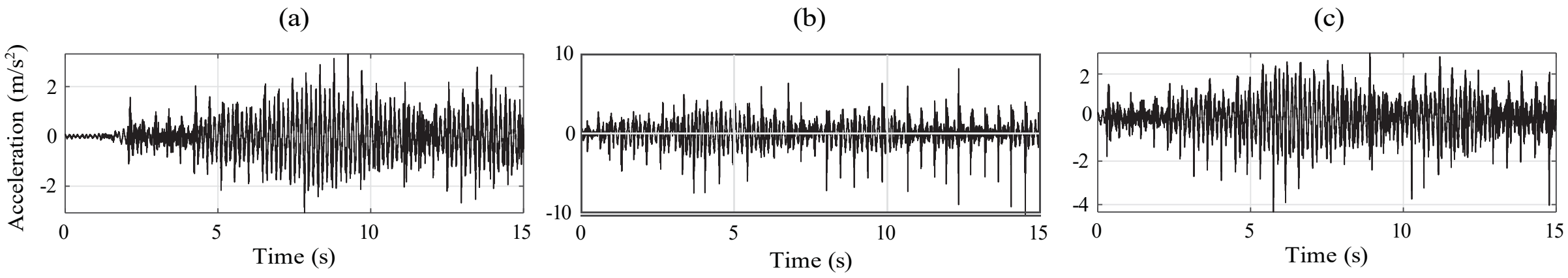

Figure 7: Acceleration time series of the sensors located at span quarters: (a) quarter; (b) mid-span; and, (c) three-quarters.

Figure 7 shows the time series recorded by the three sensors located on the bridge. Then, the FDD analysis was performed on the recordings from the three sensors. Figure 8 shows the first singular values obtained from five independent tests. It is confirmed that the first natural frequency of the bridge is $f_{b1} \approx 6.65$ Hz.

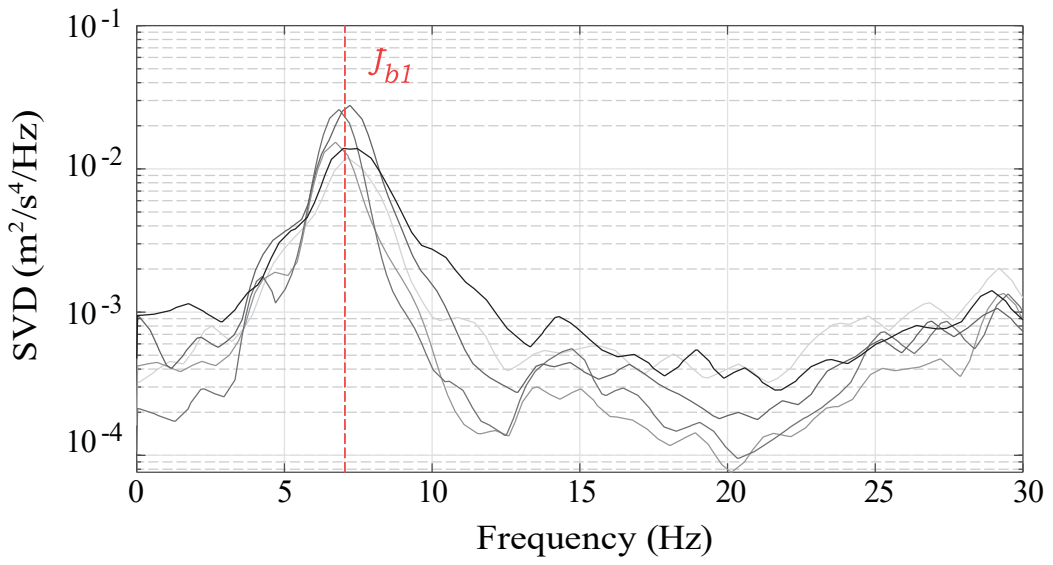

Figure 8: First singular values from five different direct inspection tests. The red dashed line represents the averaged first natural frequency of the structure.

### 4.1.2. *Indirect Sensing*

The second test consists of a drive-by inspection (see Figure 6b). To this end, 49 passes of the inspection vehicle are performed at a constant speed of 0.17 m/s. Similar to the direct sensing test, the bridge remains in operation during the test, with random pedestrian-induced excitations present. Figure 9 presents the acceleration time series recorded by the four sensors mounted on the inspection vehicle during the test.

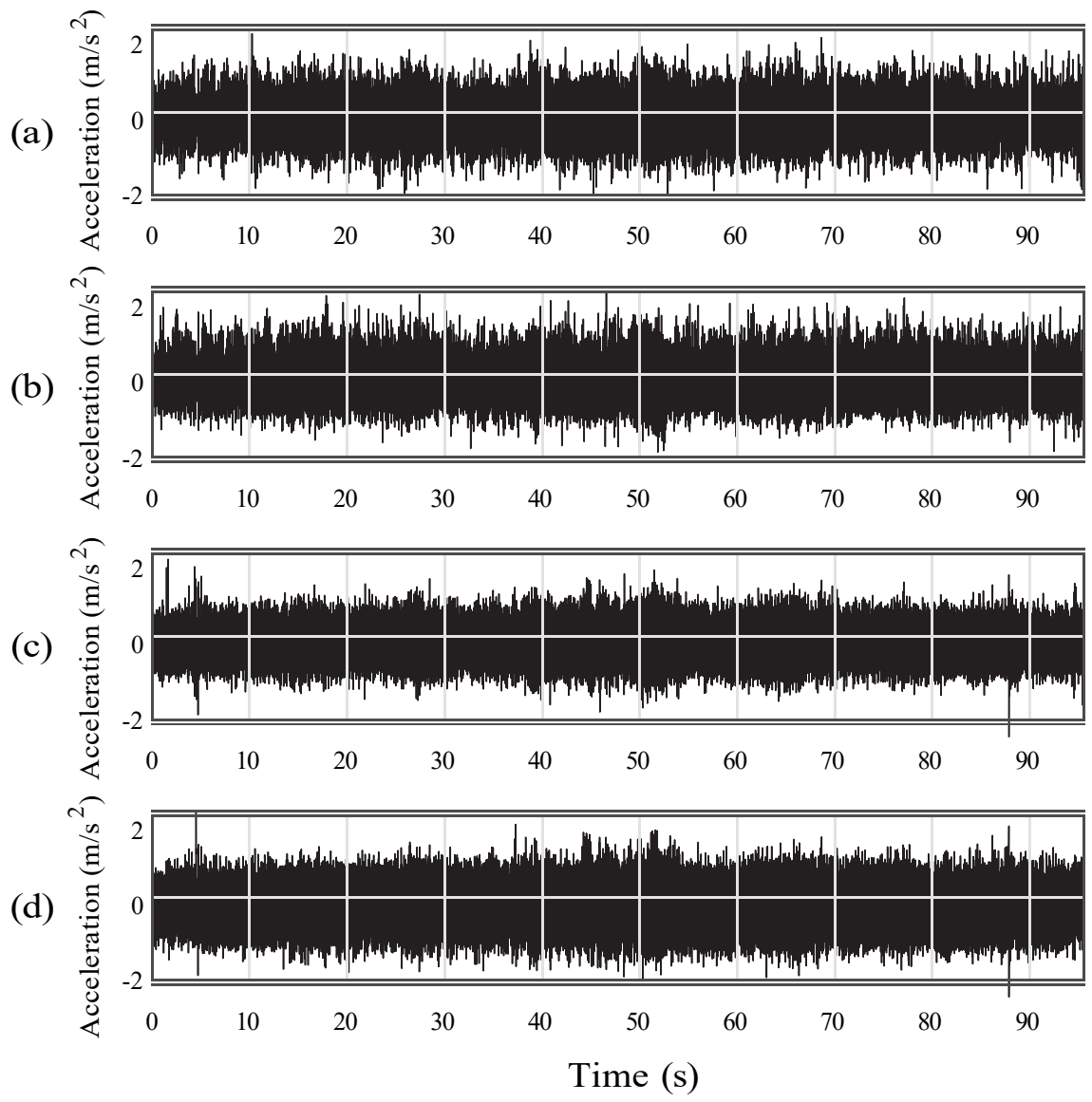

Figure 9: Acceleration time series from the sensors placed on the suspension: (a) front-right axle; (b) front-left axle; (c) rear-right axle; and (d) rear-left axle.

Figure 10 presents the first singular values obtained from the FDD considering the acceleration responses of the four mounted accelerometers. The figure displays the results from five different vehicle runs over the bridge. The average first natural frequency of the structure, identified from the peaks of the singular value plots, is $f_{b1} \approx 6.65$ Hz. Additionally, the frequencies associated with the motor in operation $f_{vd}$ of the vehicle are identified. These findings demonstrate that the drive-by bridge inspection methodology can accurately identify the first natural frequency of the structure.

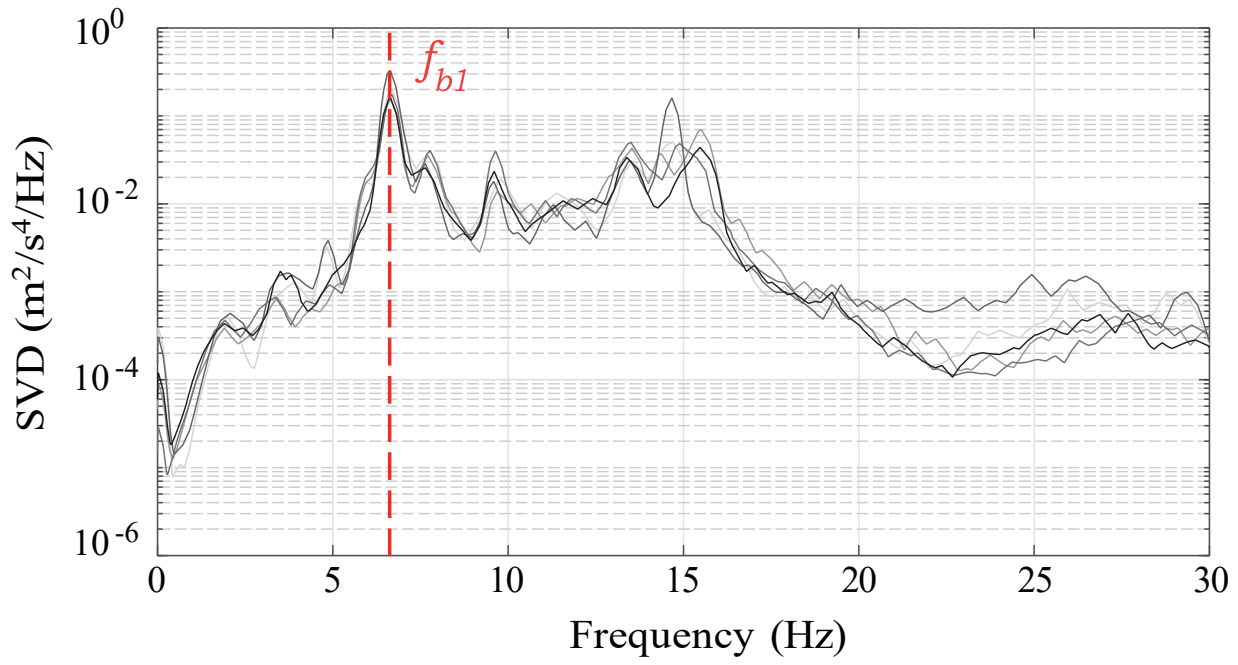

Figure 10: First singular values from five different indirect inspection tests. The red dashed line represents the averaged first natural frequency of the structure, $f_{b1}$

### *4.2. Case Study II: Bulli Colliery Bridge*

The second case study examines a real bridge located in Bulli, NSW, Australia. The bridge, named the Bulli Colliery Bridge, is a steel girder bridge that spans the Princes Highway in Bulli. The Bulli Colliery Bridge was originally built to transfer coal trucks to the Illawarra railway line, but is currently used as a pedestrian bridge. It consists of three spans, with the middle span being the focus of this study. The target span is 23.9 m long. Figure 11 presents a view of the bridge and its orientation.

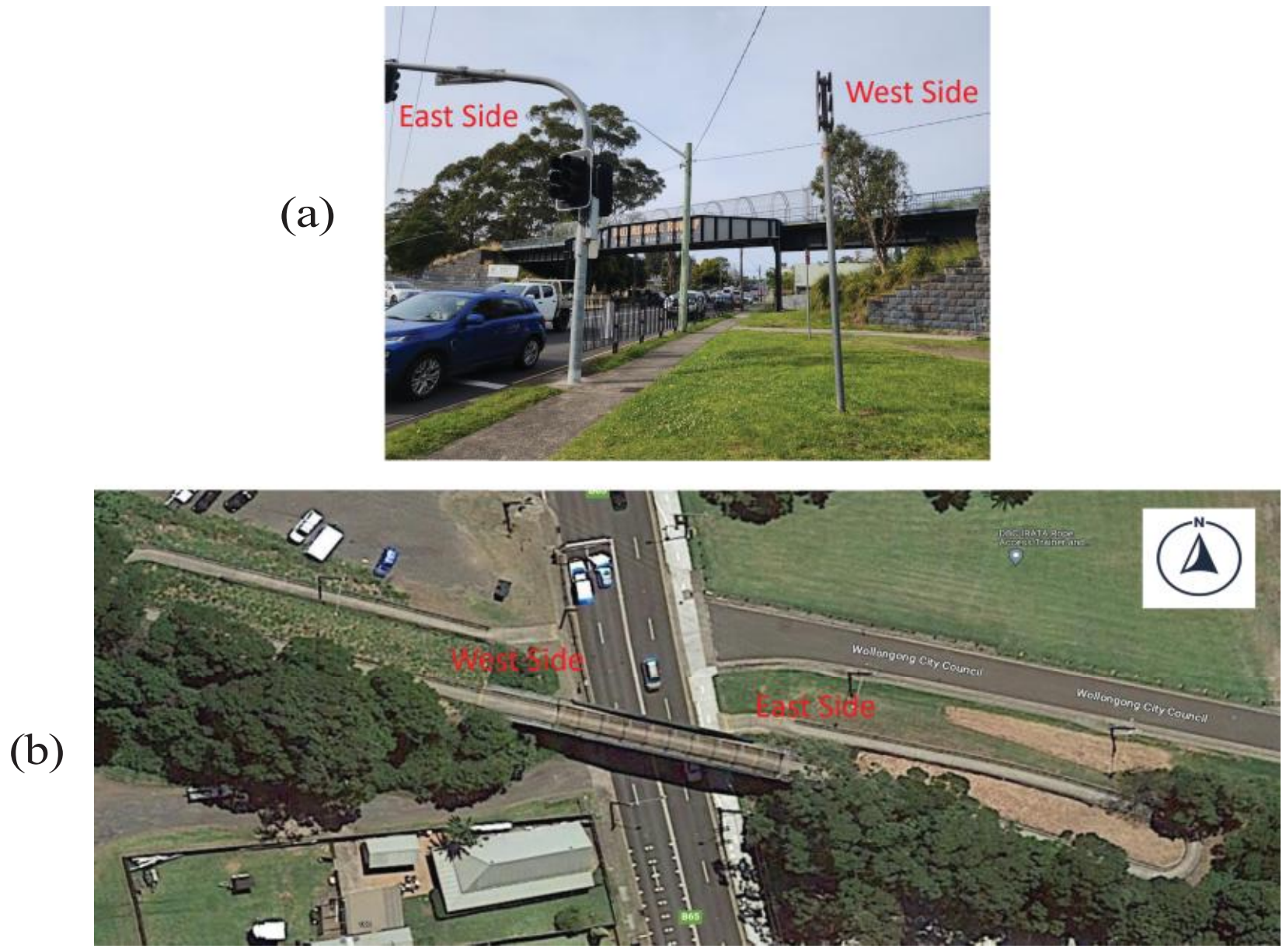

Figure 11: Bulli Colliery Bridge: (a) side view of the bridge pointing in the south direction; and, (b) top view (taken from Google Maps).

#### *4.2.1. Direct Sensing*

As in Case Study I, two tests are performed on this structure. The first test – see Figure 12a) – corresponds to a direct sensing experiment with fixed wireless sensors installed on the structure. The same sensors and setup as in the first case study (i.e., three sensors located at the quarter points of the target span) are adopted.

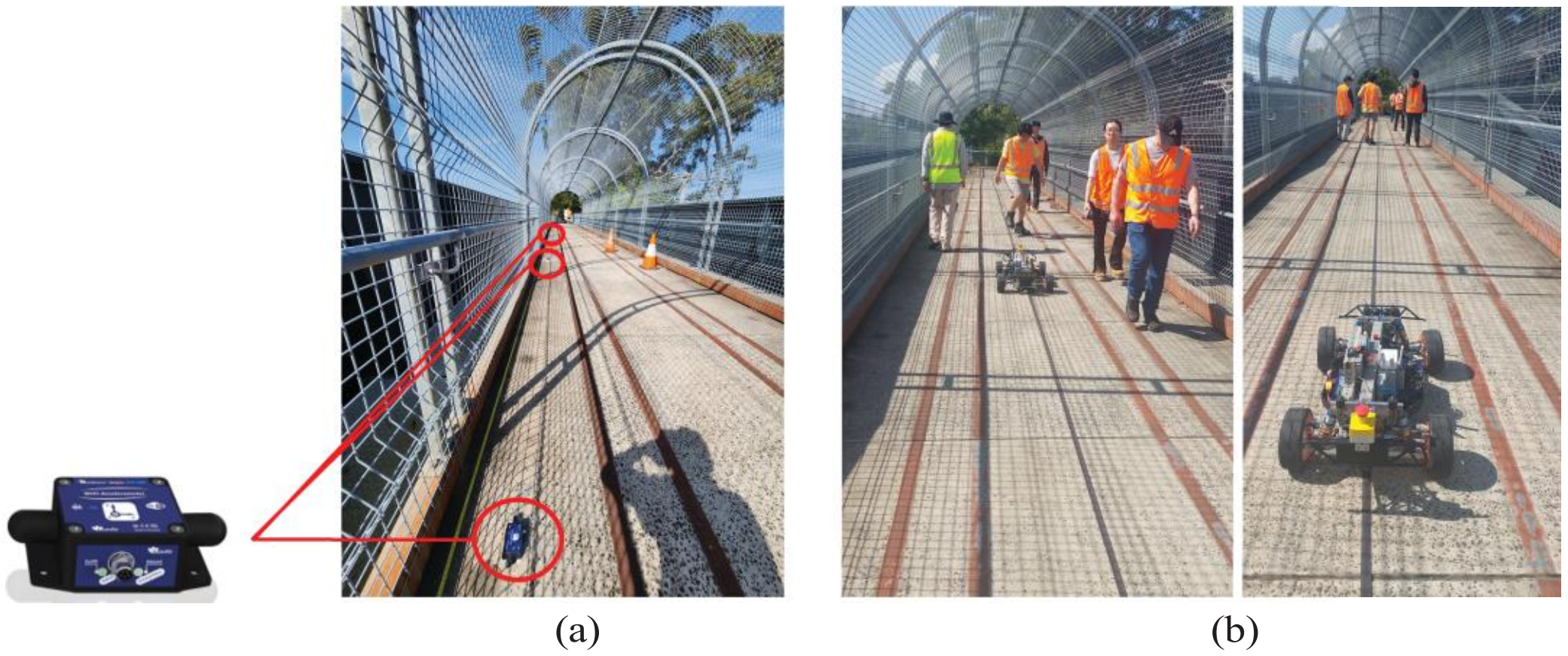

Figure 12: Bulli Colliery Bridge inspection tests: (a) direct test with three triaxial BeanAir sensors sensors directly installed on the bridge; and (b) drive-by inspection.

Figure 13 shows an example of the acceleration time series of the direct sensing test. Figure 14 presents the corresponding first singular values from five different tests obtained from the FDD, considering the three accelerometers mounted on the bridge. The first fundamental frequency of the bridge, $f_{b1}$ = 6.7 Hz, is identified and serves as a benchmark for evaluating the performance of the indirect test.

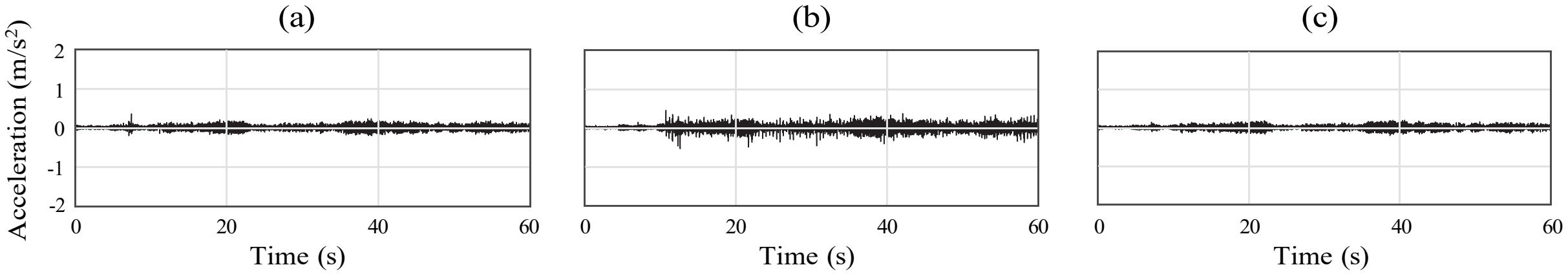

Figure 13: Acceleration time series from the sensors located at quarters of the span: (a) quarter; (c) mid-span; (b) three-quarters.

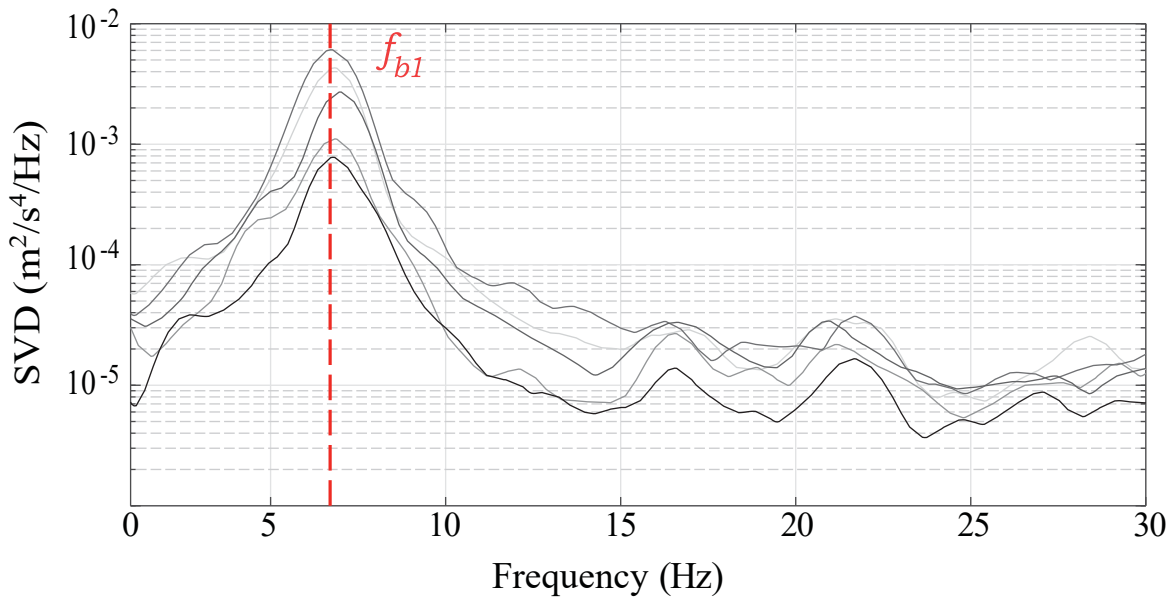

Figure 14: First singular value from five different direct inspection tests. The red dashed line represents the first natural frequency of the structure.

### 4.2.2. *Indirect Sensing*

Secondly, an indirect sensing test using drive-by bridge inspection (see Figure 12b) is conducted. The same inspection vehicle setup as in Case Study I, driving at the same speed (i.e., 0.17 m/s), is used for this purpose, and 39 crossings of the vehicle were performed. As in the first case study, the bridge remained in service during both tests, with pedestrians able to walk across it. Figure 15 shows the acceleration responses recorded by the four accelerometers on the vehicle.

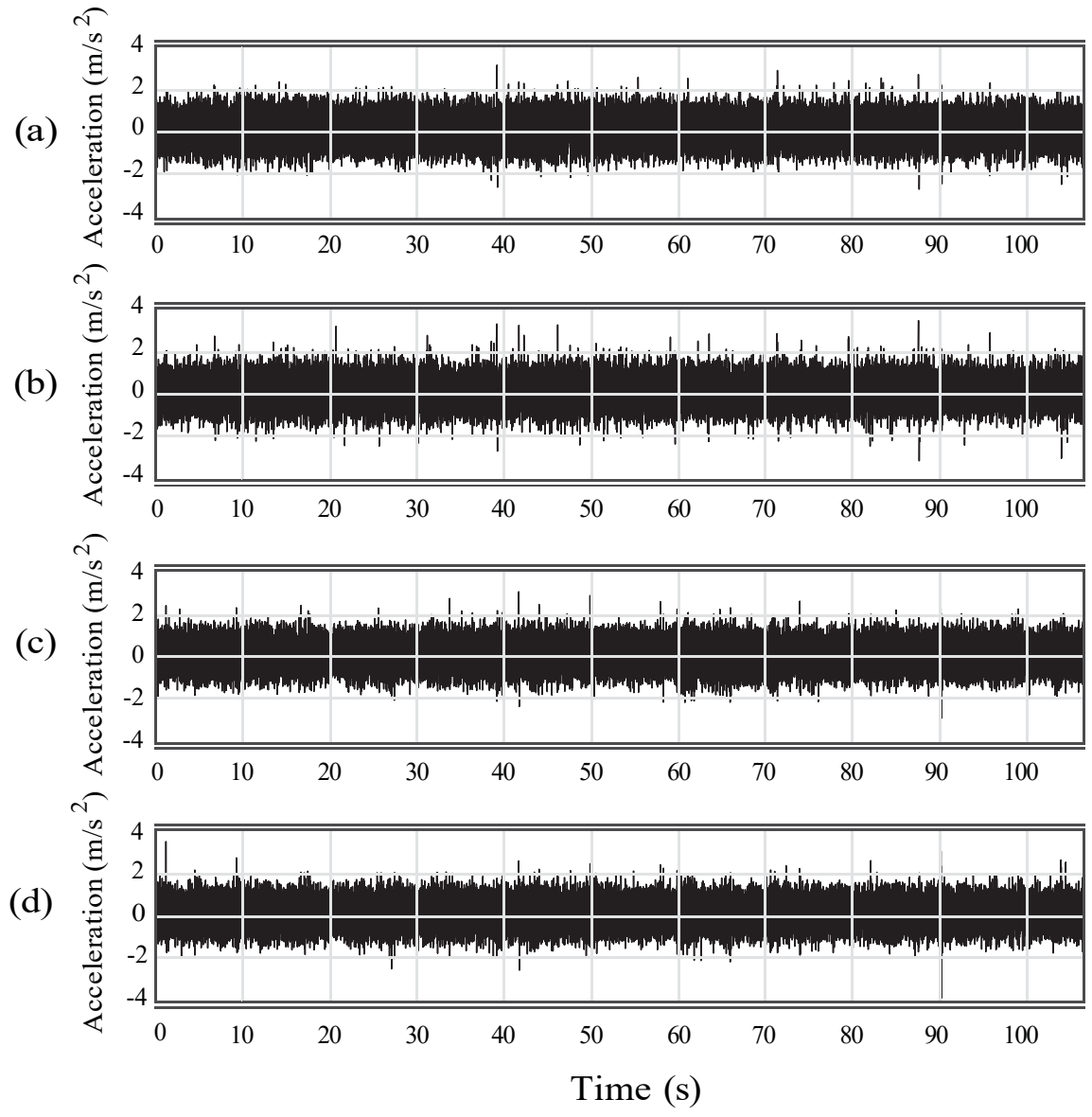

Figure 15: Acceleration time series from the sensors placed on the suspension for the indirect test performed on the Bulli Colliery Bridge: (a) front-right axle; (b) front-left axle; (c) rear-right axle; and (d) rear-left axle.

The first singular values obtained from the FDD process are presented in Figure 16 for five different test runs. As shown in Figure 16, the results confirm the successful identification of both the first natural frequency of the bridge and the driving frequency associated with the electric motor of the inspection vehicle during the indirect sensing tests. Specifically, peaks are consistently observed at approximately 6.7 Hz across all tests, corresponding to the first vibration mode of the bridge. Additionally, the frequencies associated with the operation of the electric motor, $f_{vd}$, are also identified. These findings further confirm the ability of the drive-by approach in adequately capturing the dynamic characteristics of the structure.

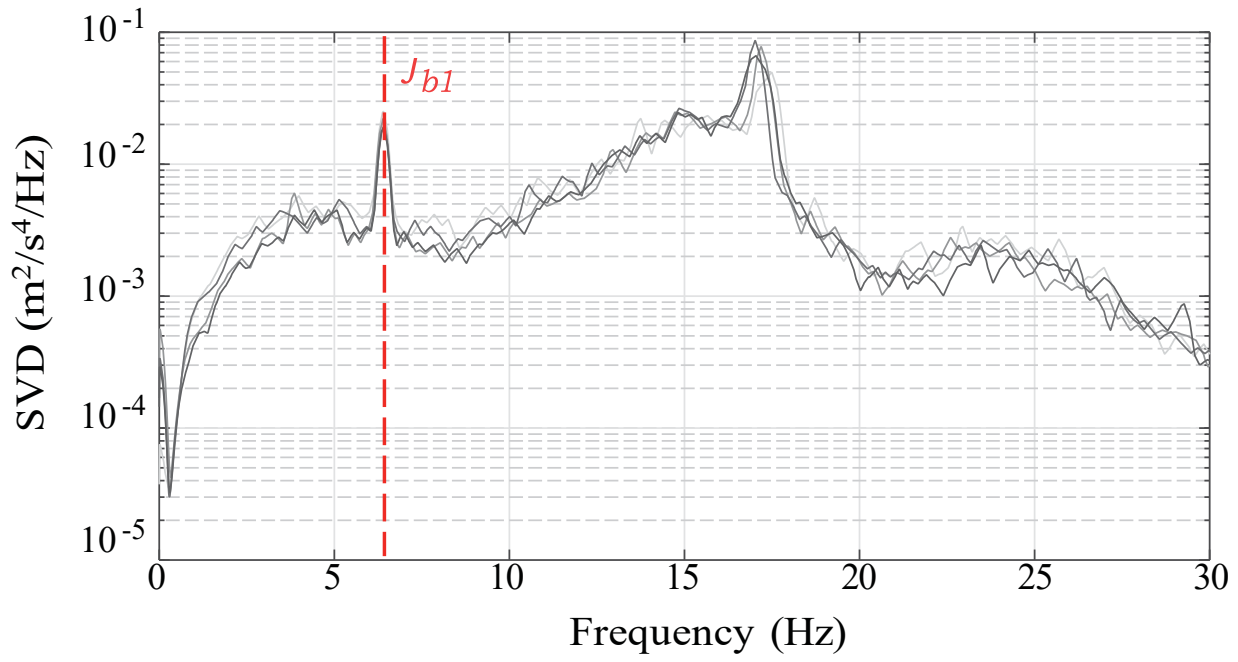

Figure 16: First singular values from five different indirect inspection tests on the Bulli Colliery Bridge. The red dashed lines represent the averaged first natural frequency of the structure $f_{b1}$

## 5. Methodology

After successfully identifying the fundamental frequency of the bridge using vehicle-mounted accelerometers – and confirming consistency with direct sensing – this section focuses further explores the potential for bridge health monitoring of drive-by sensing. To this aim, two state-of-the-art unsupervised damage assessment frameworks, previously established in [29] and [30, 31], will be used and are detailed next.

### 5.1. *Indirect SHM Using Drive-by Bridge Inspection*

This section demonstrates the potential application of the drive-by bridge inspection setup to SHM. In this stage, the data recorded by the inspection vehicle from the target bridges is processed using state-of-the-art damage assessment techniques. The first methodology, presented in [29], employs an Adversarial Autoencoder (AAE) to identify and evaluate damage. In that study, the authors compared the proposed approach with other state-of-the-art techniques, demonstrating its enhanced performance. Although the methodology was validated through a scaled experimental test, its applicability in more realistic scenarios has not yet been demonstrated.

The second methodology, presented in [30, 31], involves a time-series-based analysis using an unsupervised semantic segmentation technique based on Matrix Profile (MP). Although promising results were obtained using the matrix profile, previous work was limited to direct sensing and simplified laboratory experiments.

The current study aims to further validate the frameworks proposed in [29, 30, 31] using real data from these two bridges, specifically concerning the proposed sensing platform described earlier. In the first case study, the nominal condition of the bridge is intentionally varied by adding localised mass to simulate a variation on the bridge condition. However, in the second case study, such modification could not be performed due to the significant mass of the bridge and operational constraints, which made it impractical to introduce a measurable dynamic change. As a result, the second case study focuses solely on characterising the nominal bridge state and evaluating the robustness of the methodology to false positives.

### 5.2. *Adversarial Autoencoders (AAE)*

The first damage assessment methodology employed in this work aims to establish the baseline condition of the bridge using an unsupervised machine learning approach based on an AAE [29]. To this end, the data collected by the inspection vehicle is first processed following the procedure presented in Section 4. In addition, two pre-processing steps are applied to the first singular values $\mathbf{S}_1$, as presented in Equation 2. First, the singular values from the CPSD of the acceleration recordings of the vehicle are filtered within the frequency range of 1 to 10 Hz, which encompasses the first natural frequency of the structure, as discussed in Sections 4. The filtered signal, denoted as $\mathbf{S}_{1,filt}$, is used in the subsequent steps.

Then, sets of three $\mathbf{S}_{1,filt}$ from random vehicle crossings are averaged and normalised to the range of $[0, 1]$ using a *min-max* scaling approach, as follows:

$$\mathbf{S} = \frac{\mathbf{S}_{1,filt} - \min(\mathbf{S}_{1,filt})}{\max(\mathbf{S}_{1,filt}) - \min(\mathbf{S}_{1,filt})} \quad (3)$$

where, $\mathbf{S}$ denotes the averaged, normalised and filtered first singular values. Normalisation facilitates faster convergence of the training process. The resulting processed data, $\mathbf{S}$, is subsequently used as an input for training the AAE employed in this study. Figure 17 illustrates the first methodology as a potential application for ISHM, and it is explained next.

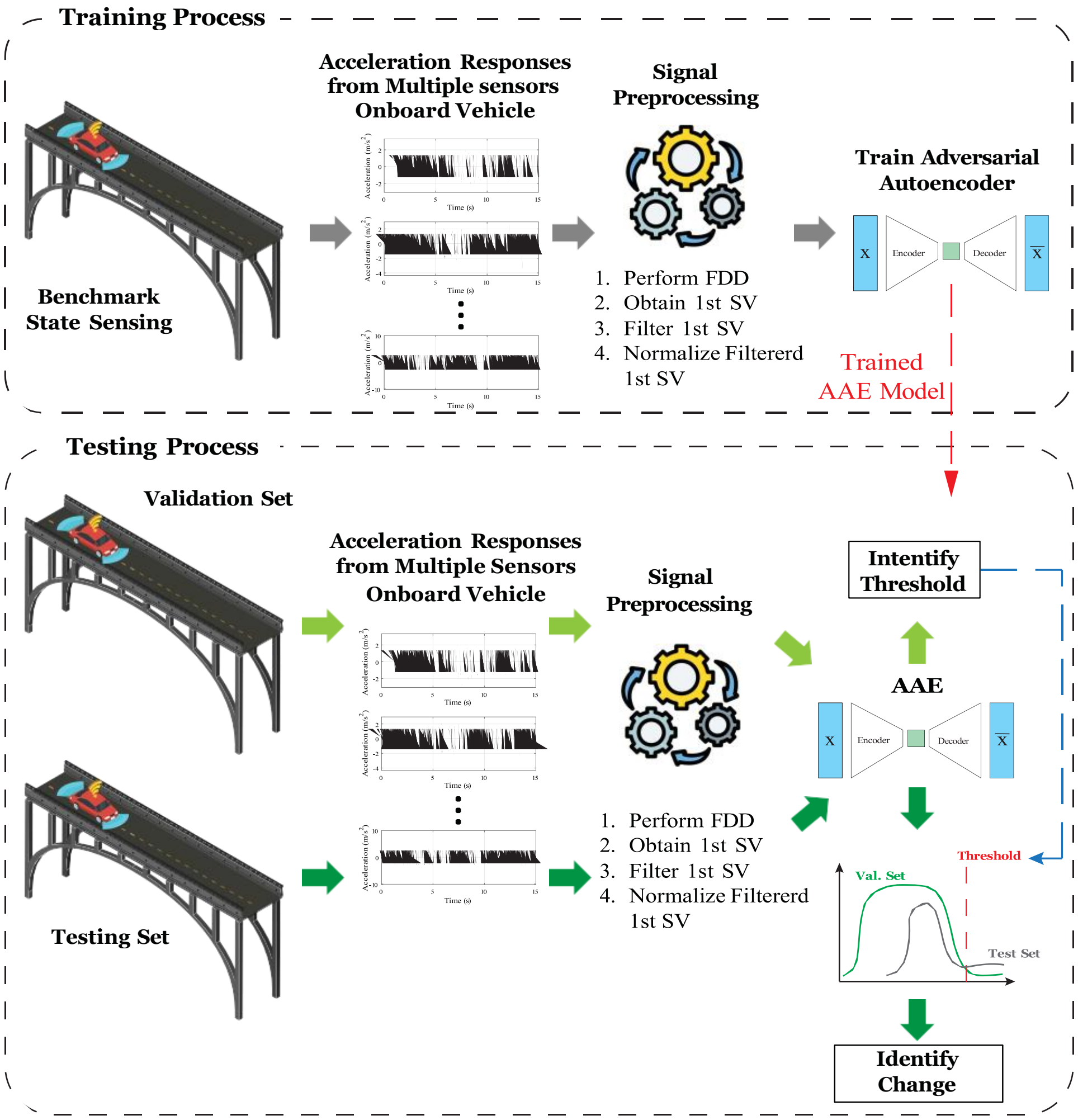

Figure 17: The framework based on AAE to prove the potential application of drive-by bridge inspection in SHM [29].

AAE extend traditional autoencoders by incorporating adversarial training to impose a desired prior distribution on the latent space, thereby improving the structure and continuity of the encoded representation [33]. An AAE consists of three components: an *encoder*, a *decoder*, and a *discriminator* network. While the encoder-decoder function as in a standard autoencoder for data reconstruction, the discriminator encourages the latent codes to match a predefined prior distribution (e.g., Gaussian distribution) [34]. Figure 18 shows a typical AAE structure.

Let the input data vector be denoted as $x$, and the corresponding reconstructed output as $\overline{x}$. The encoder maps $x$ to a latent code $y$ with encoding distribution $q(y|x)$, and the decoder reconstructs $\overline{x}$ from $y$. The latent space is regularised by aligning the *aggregated*

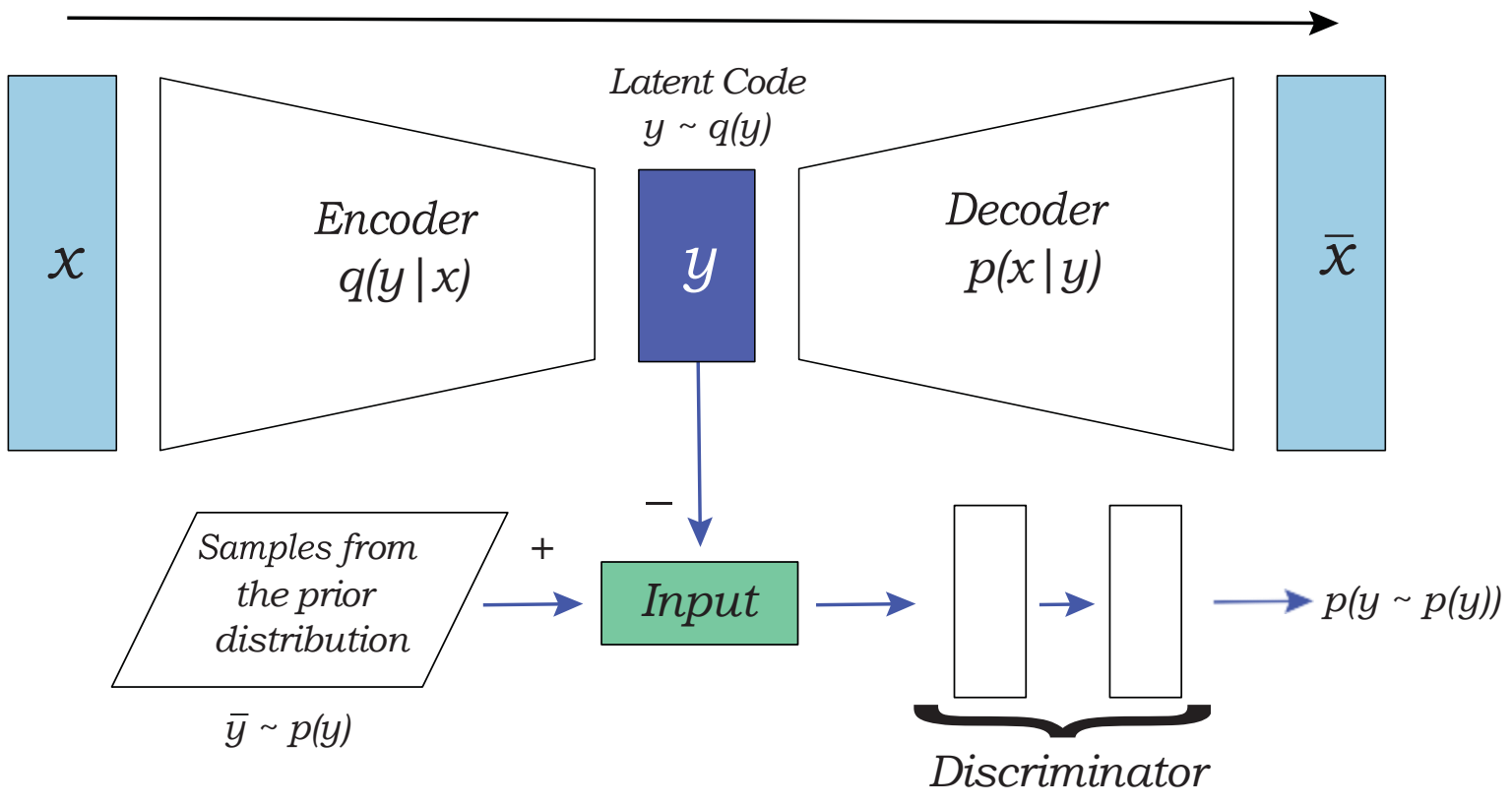

Figure 18: A typical adversarial autoencoder structure [35].

*posterior* $q(y)$ with a chosen prior $p(y)$, i.e.:

$$q(y) = \int_x q(y|x)p_D(x)\, dx, \tag{4}$$

where $p_D(x)$ is the empirical data distribution. The discriminator network receives samples from both $q(y)$ and $p(y)$, learning to classify whether a sample comes from the prior or from the encoder. This adversarial setup involves two alternating phases during training: (1) the *reconstruction phase* where the autoencoder minimises the reconstruction error between $x$ and $\bar{x}$, and (2) the *regularisation phase* where the discriminator tries to distinguish $y \sim q(y)$ from $y_{\text{true}} \sim p(y)$, while the encoder is trained to fool the discriminator. The *discriminator loss* is defined as:

$$Loss_D = -\frac{1}{n}\sum_n \left[\log(D(y_{\text{true}})) + \log(1 - D(y))\right], \tag{5}$$

where $D(\cdot)$ is the discriminator output, $y_{\text{true}} \sim p(y)$, $y \sim q(y)$, and $n$ is the number of training samples. The discriminator has a sigmoid output layer, producing values in the range of (0, 1), corresponding to the probability of a sample being drawn from the prior distribution. The *generator loss*, used to update the encoder, combines the reconstruction error and the adversarial objective is given by:

$$Loss_G = \frac{1}{n}\sum_n (x - \bar{x})^2 - \frac{1}{n}\sum_n \log(D(y)). \tag{6}$$

In this work, $x$ is represented by the preprocessed signals, $\mathbf{S}$. Furthermore, the hyperparameters used in this work correspond to the ones used in [29].

#### *5.2.1. Training and Testing the Model:*

As previously mentioned, the aim of implementing a damage assessment methodology in this work is to demonstrate the potential of drive-by bridge inspection for identifying structural damage. However, due to the challenges of modelling real damage in in-service structures, this study adopts two complementary approaches. In the first case study (i.e., the UNSW bridge), bridge condition variation is simulated by introducing added mass to the structure. In the other case study (i.e., Bulli bridge), the focus is on detecting false positives using only responses from the healthy, or nominal, condition of the structure. To this end, the preprocessed acceleration signals recorded by the instrumented inspection vehicle (i.e., **S**) are used to train and test the AAE model. These signals presented in Section 4, representing the nominal state of the structure, are split into training and testing sets in an 80:20 ratio. During model validation, a damage threshold is defined as the 90$^{th}$ percentile of the reconstruction error $E_r$ – see also Figure 17:

$$E_r = (x - \bar{x})^2, \tag{7}$$

where the thresold follows the recommendations presented in [35].

Once the damage threshold is defined, the testing set is passed through the AAE network, and the corresponding reconstruction errors are computed. These errors are then compared against the predefined threshold to identify if damage is present.

To demonstrate the methodology's potential for ISHM using drive-by inspection, it is crucial to evaluate both its ability to minimise false positives (i.e., healthy samples misclassified as damaged) and to identify true negatives consistently. A low false positive rate suggests that the methodology reliably characterises the nominal state of the structure, which is a necessary foundation for any SHM system. However, robustness to false positives alone does not guarantee sensitivity to actual damage. Therefore, complementary validation, such as a separate test on the UNSW bridge where damage is simulated by adding localized mass, is essential to assess the method's capacity to detect real structural changes.

### *5.3. Matrix Profile (MP)*

The second approach is based on the framework presented in [30, 31]. In drive-by inspection, the amount of available data is often limited due to constraints in testing opportunities or operational conditions. This presents a challenge for the AAE approach, which relies on large training datasets. One potential solution is the MP method [36], which can detect change points in bridge states using limited observations. Therefore, this study investigates the use of this technique as a second approach.

Figure 19 provides a schematic overview of the MP-based framework. The MP method is typically used in combination with the Corrected Arc Crossings (CAC) to detect potential change points in the bridge state. If the bridge remains in a consistent condition throughout the observation period, the algorithm is not expected to detect any change points. This method has been validated in [30, 31] using numerical simulations and laboratory experiments, but it has not yet been tested on full-scale bridges.

#### *5.3.1. Change Point Detection*

In the change point detection framework, drive-by inspections are carried out periodically over a defined inspection period. During each inspection, vehicle responses are recorded as the vehicle passes over the bridge. The responses from multiple inspections are then combined to form a single observation sequence. The framework analyses this sequence to identify potential change points in the bridge state.

Let $\mathbf{G} = [\{G_1(\tau)\}, \{G_2(\tau)\}, \ldots, \{G_n(\tau)\}]$ , denote the set of responses collected from $n$ independent passages of test vehicle over the bridge. Each $G_i(\tau)$ is a vector of length $m$, representing the response recorded during the $i$-th passage, where $\tau \in \{t_1 : t_m\}$ denotes the discrete time index [30]. The total length of the observation sequence $\mathbf{G}$ is defined as $p = m \times n$. It is assumed that a change in bridge condition happens at an unknown time $t_c \in \{t_1 : t_p\}$. The subsequence $\{\hat{G}_k\}_l \in \mathbb{R}^l$ is defined as a continuous subset of the values in $\mathbf{G}$ with length $l$. The starting position of the subsequence is denoted by $k$, where $k \in \{1 : p-l+1\}$. The subsequence can be represented as $\{\hat{G}_k\}_l = \{\mathbf{G}_k, \mathbf{G}_{k+1}, \dots, \mathbf{G}_{k+l+1}\}$, in which $\mathbf{G}_k$ denotes the $k$-th element of $\mathbf{G}$. $\{\hat{G}_k\}_l$ can be treated as a sliding window applied to $\mathbf{G}$, through which all possible subsequences of length $l$ are obtained. This results could be collected in $\mathbf{L}_l = \left[\{\hat{G}_1\}_l, \{\hat{G}_2\}_l, \dots, \{\hat{G}_{p-l+1}\}_l\right]$, where the $k$-th subsequence $\{\hat{G}_k\}_l$ is donated as $\mathbf{L}_l^k$.

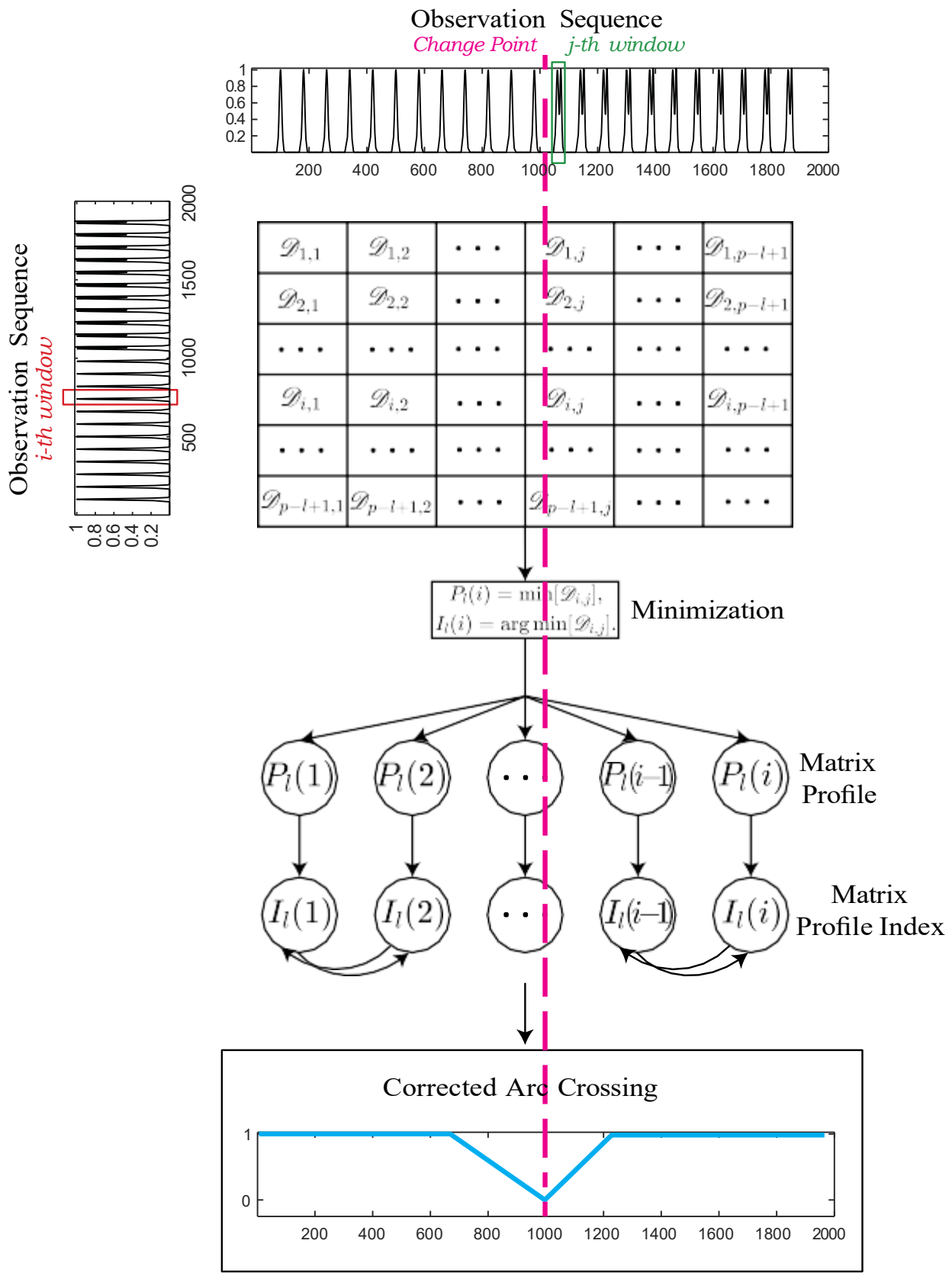

Figure 19: Overview of the MP method.

Once the full set of subsequences is constructed, the all-pairs similarity search is performed to identify the first-nearest neighbour of each subsequence. During the search procedure, trivial matches, such as self-matching ones, are excluded. For each pair of subsequences $i$ and $j$ ($\forall i, j \in \{1 : p - l + 1\}$) the similarity is measured using the $z$-normalised Euclidean distance [37]. The distance $D_{i,j}$ for two subsequences $\mathbf{L}_l^i$ and $\mathbf{L}_l^j$ could be calculated asL:

$$D_{i,j} = \sqrt{\sum_{r=1}^{l} \left[ \mathbf{Z}\left(\mathbf{L}_l^i\right)_r - \mathbf{Z}\left(\mathbf{L}_l^j\right)_r \right]^2}, \tag{8}$$

where $\left(\mathbf{L}_l^i\right)_r$ represent the $r$-th element in the subsequence $\mathbf{L}_l^i$ and $\mathbf{Z}(\cdot)$ is the $z$-score normalisation operator. Considering the properties of the z-score, the distance expression can be simplified as:

$$\mathrm{D}_{i,j} = \sqrt{2\left( l - \frac{\sum_{r=1}^{l} \left( \mathbf{L}_l^j \cdot \mathbf{Z}(\mathbf{L}_l^i) \right)_r}{\sigma_j} \right)}. \tag{9}$$

Here, the $\sigma_j$ is the standard deviation of subsequence $\mathbf{L}_l^j$. The computational complexity of Equation 9 can be reduced by applying the convolution theorem [30]. Based on the calculated distances, the matrix profile $P_l \in \mathbb{R}^{p-l+1}$ is constructed as a vector, where each element represents the minimum distance between a given subsequence and its nearest neighbour. The position of this nearest neighbour is recorded in the corresponding MP index $I_l \in \mathbb{Z}^{p-l+1}$ [38],

$$P_l(i) = \min[\mathrm{D}_{i,j}], \quad \forall j \in \{1 : p - l + 1\}, \tag{10}$$

$$I_l(i) = \arg\min[\mathrm{D}_{i,j}], \quad \forall j \in \{1 : p - l + 1\}. \tag{11}$$

Since the total number of subsequences is $p - l + 1$, the searching process is performed for each subsequence individually. The resulting MP and its index are $p - l + 1$ dimensional vectors. These results will be used for the next step of change point detection.

Each entry in the MP index defines a pair $(i, j)$, where $j$ is the nearest neighbour of subsequence $i$. This pair can be represented as an arc from location $i$ to location $j$. The arc curve (AC) can be defined as a vector in which each element indicates the number of arcs that cross over the corresponding location $i$ [39]. When the condition of the system changes at time $t_c$, most subsequences are expected to find their nearest neighbours within the host regime. Since only a small number of arcs cross $t_c$, the AC reaches the lowest value at this point. Low AC values can also be observed near the beginning and end of the series. This is due to the limited number of arcs that can cross at this boundary area. To solve this edge effect, the AC is normalised with respect to an Idealised Arc Curve (IAC), resulting in the CAC:

$$CAC(i) = \min\left\{ \frac{AC(i)}{IAC(i)}, 1 \right\}. \tag{12}$$

The IAC denotes the number of arcs obtained if all matches were uniformly sampled from the entire time series. This idealised curve forms an inverted parabola, peaking at $1/(2p)$ [30]. The resulting CAC is a vector of length $p - l + 1$, consistent with the length of the MP and its index. The range of CAC is located in [0, 1], and the minimum value indicates the most likely location of a change point in the series.

In the context of this work, the MP-based framework is applied to vehicle responses collected by the customized inspection vehicle. The recorded vehicle responses are processed using the methodology outlined in Section 4 to extract the first singular value from the CPSD. The extracted singular values are assembled sequentially to form the observation sequence **G**. Change point detection within this sequence indicates potential changes in bridge condition during the inspection period.

## 6. Results

### 6.1. Case Study I:

#### 6.1.1. Bridge Condition Assessment using AAE

Following the procedure described in Section 5.2, this section demonstrates the potential application of the AAE-based methodology for drive-by bridge inspection. First, 49 vehicle crossings over the bridge were analysed. From these crossings, the first singular values of the CPSD from the four acceleration responses of **S** (see Equation 3) were extracted and preprocessed as described in Section 5.2. These samples were then divided into training and testing sets, and the AAE was subsequently trained. Once the AAE is trained, the threshold is identified as 0.023. Figure 20 shows an example of the reconstruction of a **S** sample after training, showing the potential of the AAE to reconstruct the nominal state of the structure.

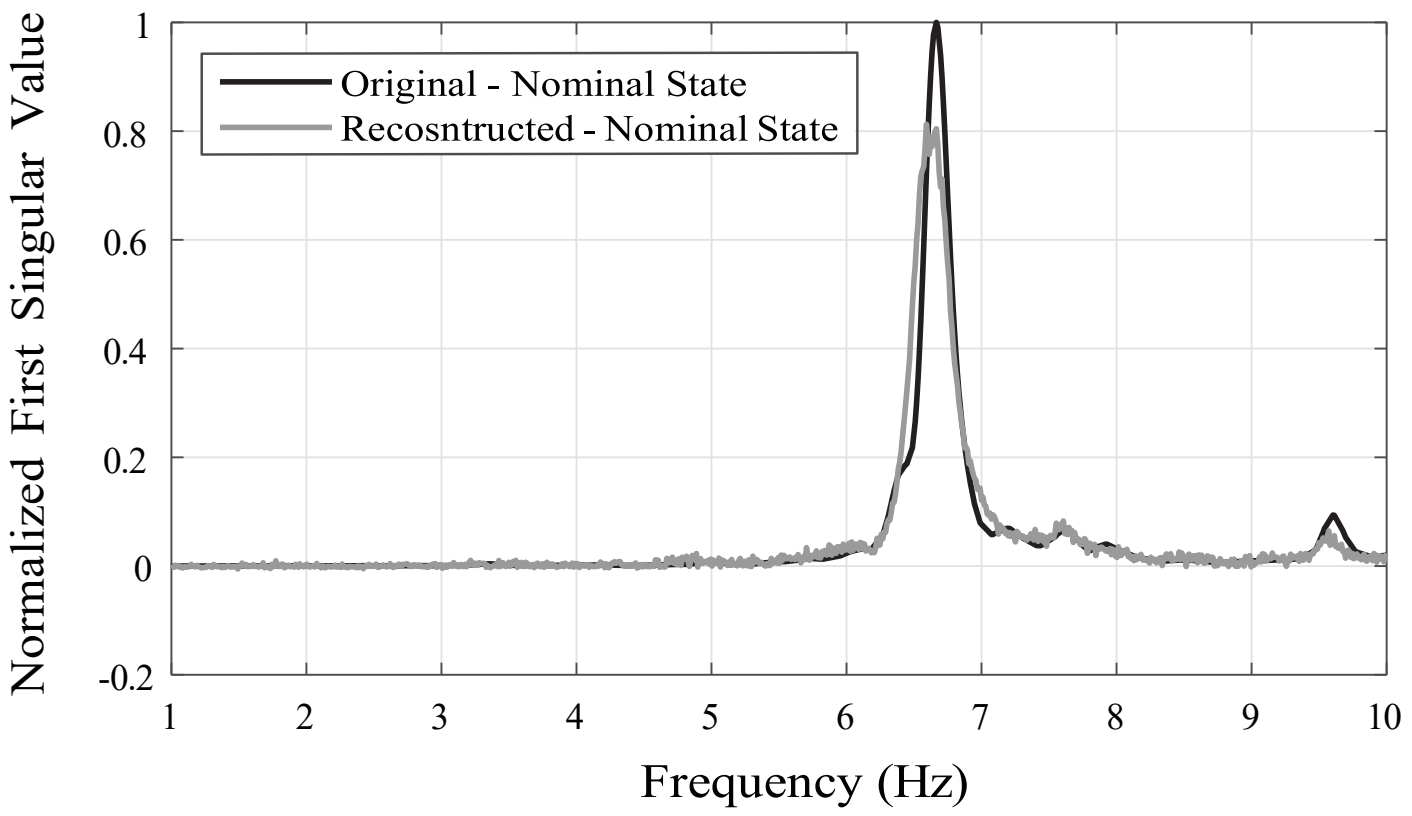

Figure 20: Reconstruction of a **S** sample using the presented AAE.

To test the bridge condition assessment approach, damage was simulated by adding mass to the bridge (i.e., five people standing at mid-span), as shown in Figure 21. The vehicle crossed the modified bridge state ten times, and the corresponding acceleration responses were pre-processed following the methodology described in Section 5.2. These pre-processed signals were then fed into the trained AAE, and the resulting reconstruction error was used to demonstrate the ability to detect structural variations. Figure 22 shows an example of the reconstruction of the frequency spectrum from the modified state of the bridge, from which it is clear that the AAE fails to reconstruct the sample from the varied bridge state.

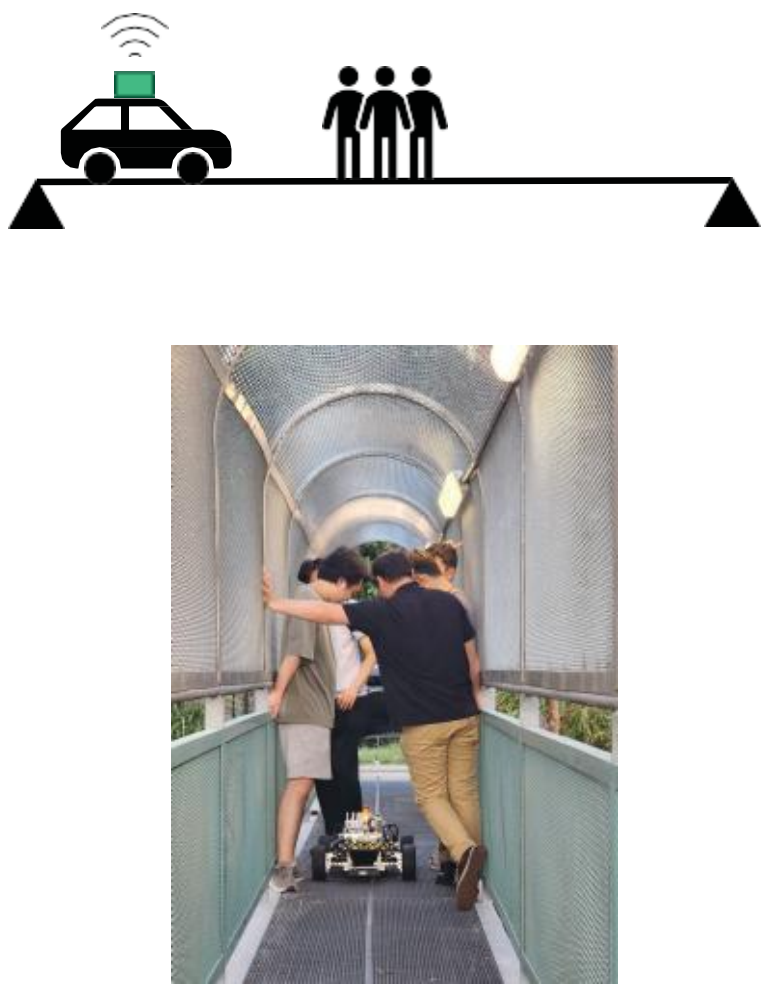

Figure 21: Drive-by bridge inspection with simulated damage.

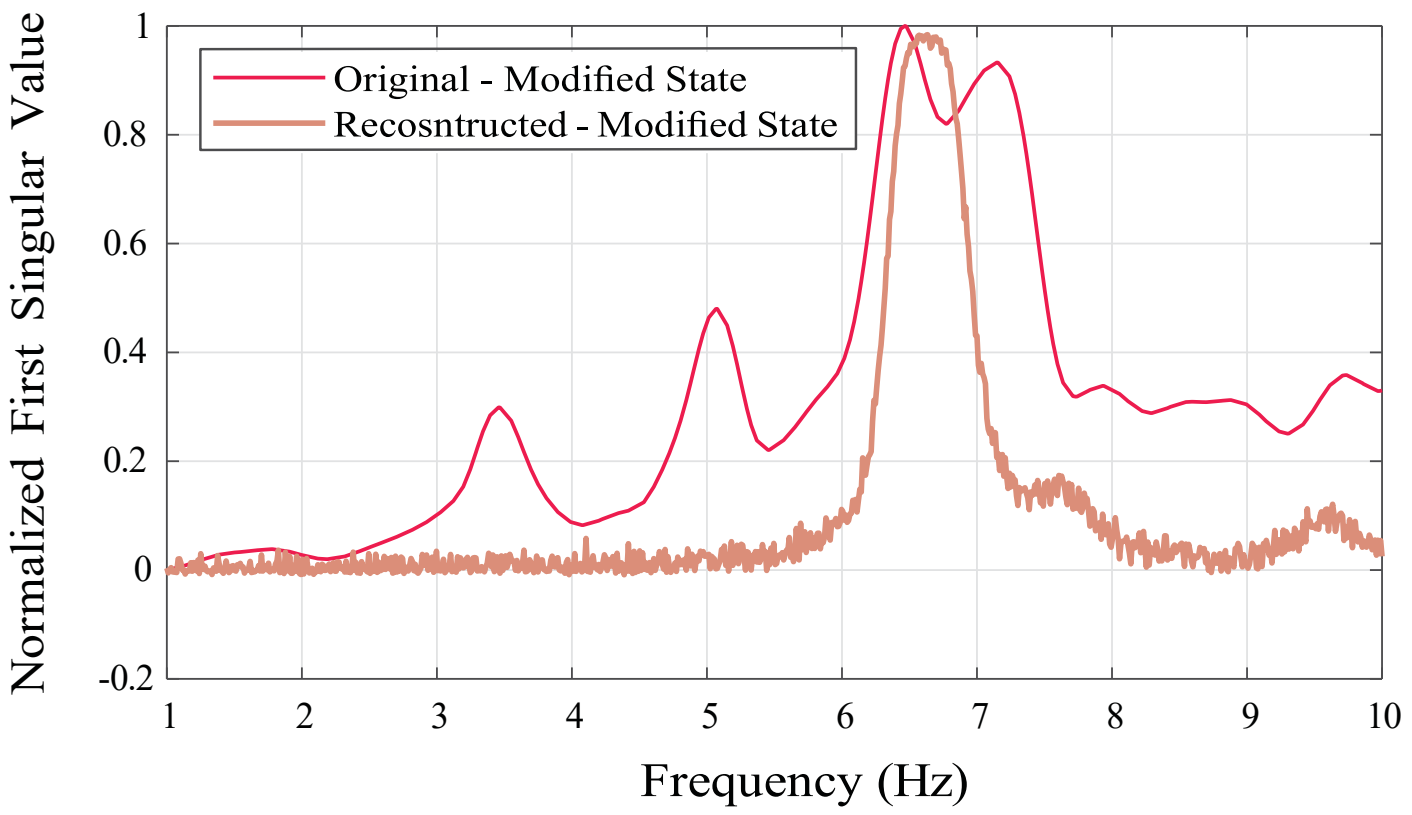

Figure 22: Reconstruction of a **S** sample from the modified state of the bridge using the presented AAE.

Figure 23 presents the reconstruction error obtained during the testing phase using the AAE-based methodology. It is evident that the reconstruction error for the simulated damage samples (i.e., modified bridge state with added mass) is significantly higher than that for the nominal state of the structure, indicating a variation in the bridge condition. It should be noticed that all samples from the nominal bridge state were correctly classified as healthy, resulting in no false positives. Likewise, all samples from the modified bridge state were correctly identified as damaged, yielding no false negatives. These results demonstrate a perfect separation between true negatives and true positives, highlighting the potential of the proposed methodology for accurately assessing bridge condition through drive-by inspections.

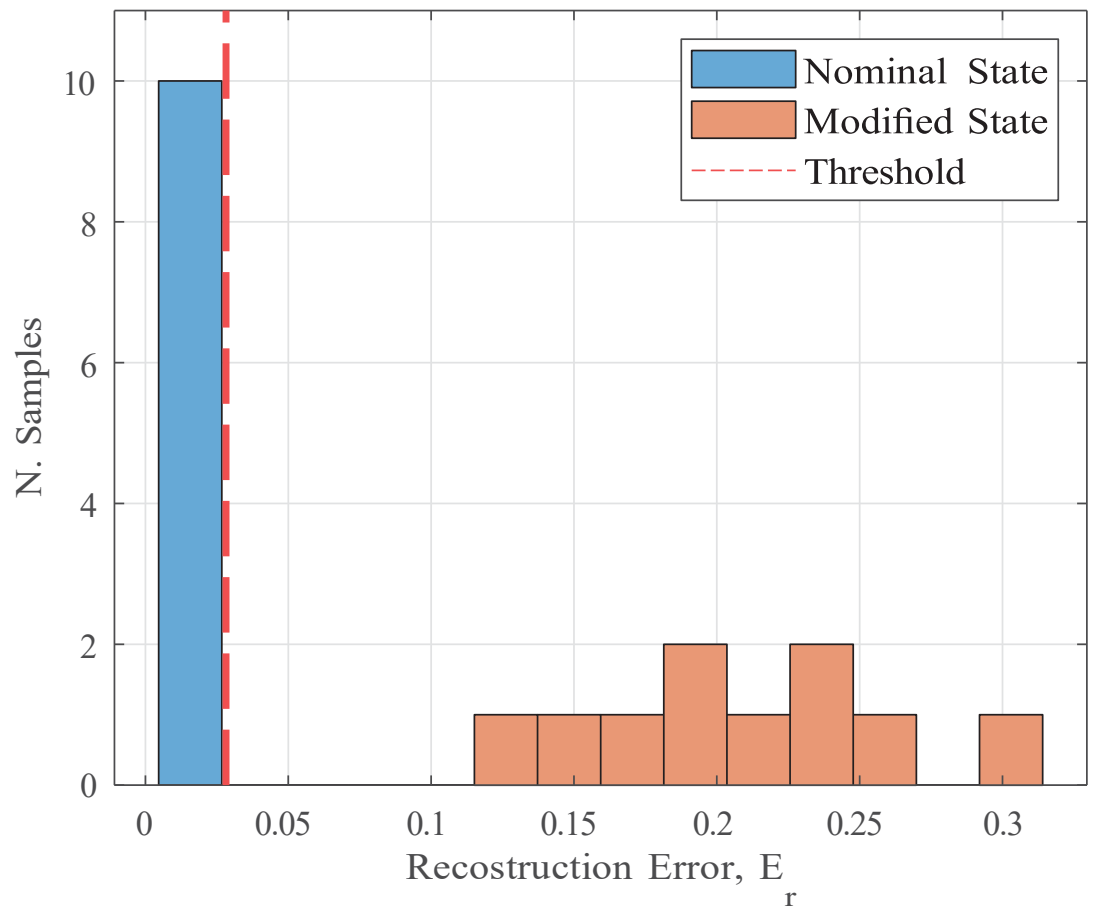

Figure 23: Reconstruction errors from healthy and simulated damage samples.

Finally, it is worth noting that, as reported in [29], the AAE-based methodology is only capable of identifying damage and assessing its severity, based on the direct relationship between the reconstruction error and the extent of the damage. Identifying the type, source, and location of the damage is beyond the scope of the presented methodology.

### *6.1.2. Bridge Condition Assessment using MP*

This section examines the performance of the proposed MP method using experimental data obtained from the pedestrian bridge located at UNSW. As outlined in Section 4.1, 49 vehicle passages were conducted, and the first singular value of the CPSD matrix was extracted for each run.

First, a case is studied in which every sample in the observation sequence corresponds to the benchmark state. The aim of this investigation is to evaluate the robustness of MP to false positives. A new singular value sample is generated by averaging five randomly selected samples out of 49. This new sample consists of 900 points or spectral lines, covering the frequency range from 1 to 10 Hz, and is normalised to the range of $[0, 1]$. For the construction of observation sequences, a total of 30 samples are produced through this process. Therefore, the size of the observation sequence will be 27,000 data points, e.g. 900 (spectral lines) $\times$ 30 (samples). The resulting observation sequence is shown in the first row of Figure 24, and the corresponding CAC profile is presented in the second row. The resulting CAC remains close to a constant value of 1 throughout the entire sequence, indicating that the method does not detect any change points.

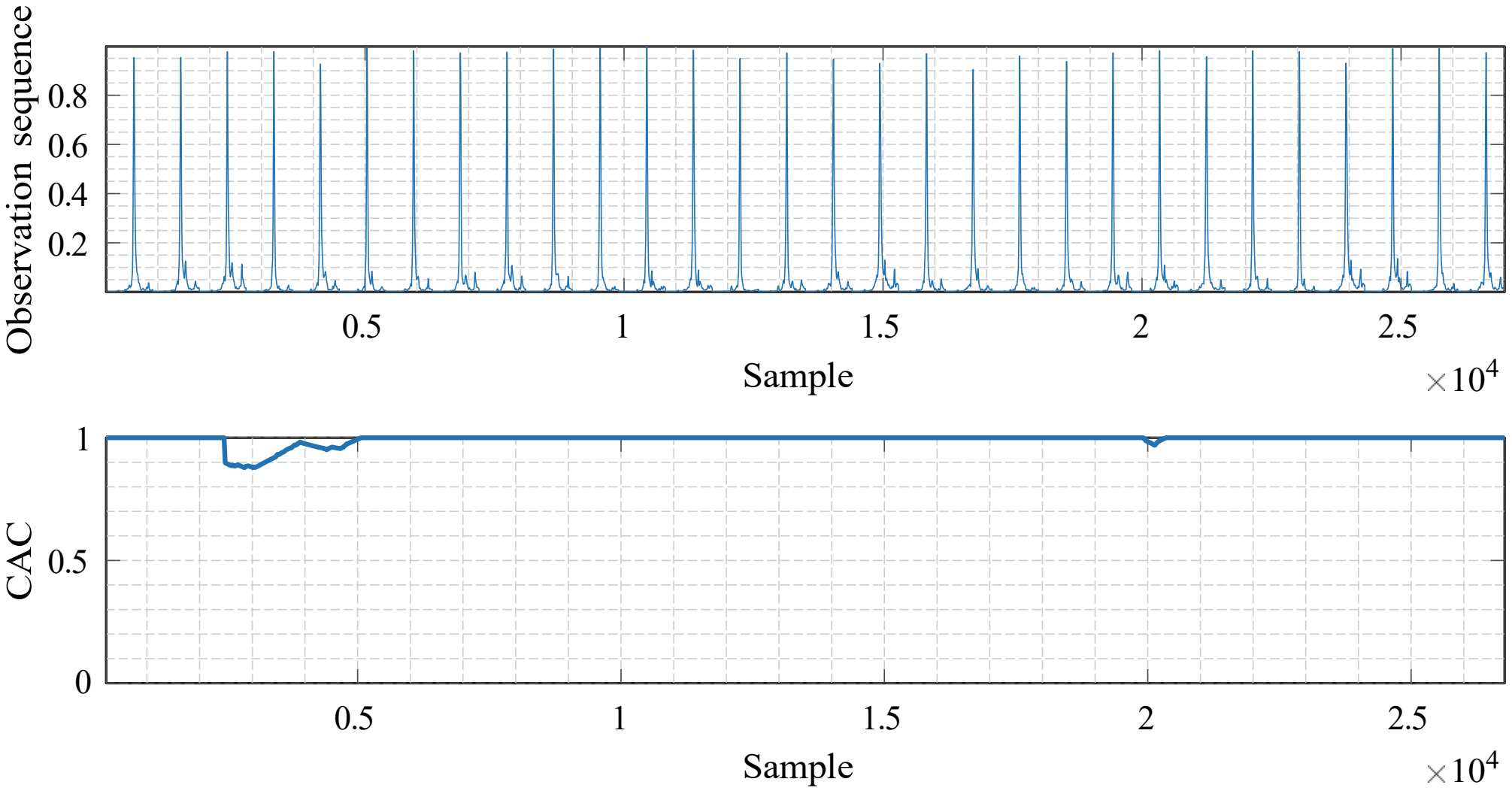

Figure 24: Observation sequence and corresponding CAC profile for the nominal state of UNSW Bridge.

An extended assessment is performed by generating 1,000 independent observation sequences under the benchmark state. The CAC is computed for each sequence, and the results are shown as a 3D surface in Figure 25. As illustrated, the surface remains close to the CAC value of one throughout all observation sequences, with only minor variations. This result confirms that the method produces stable outputs for the benchmark bridge condition.

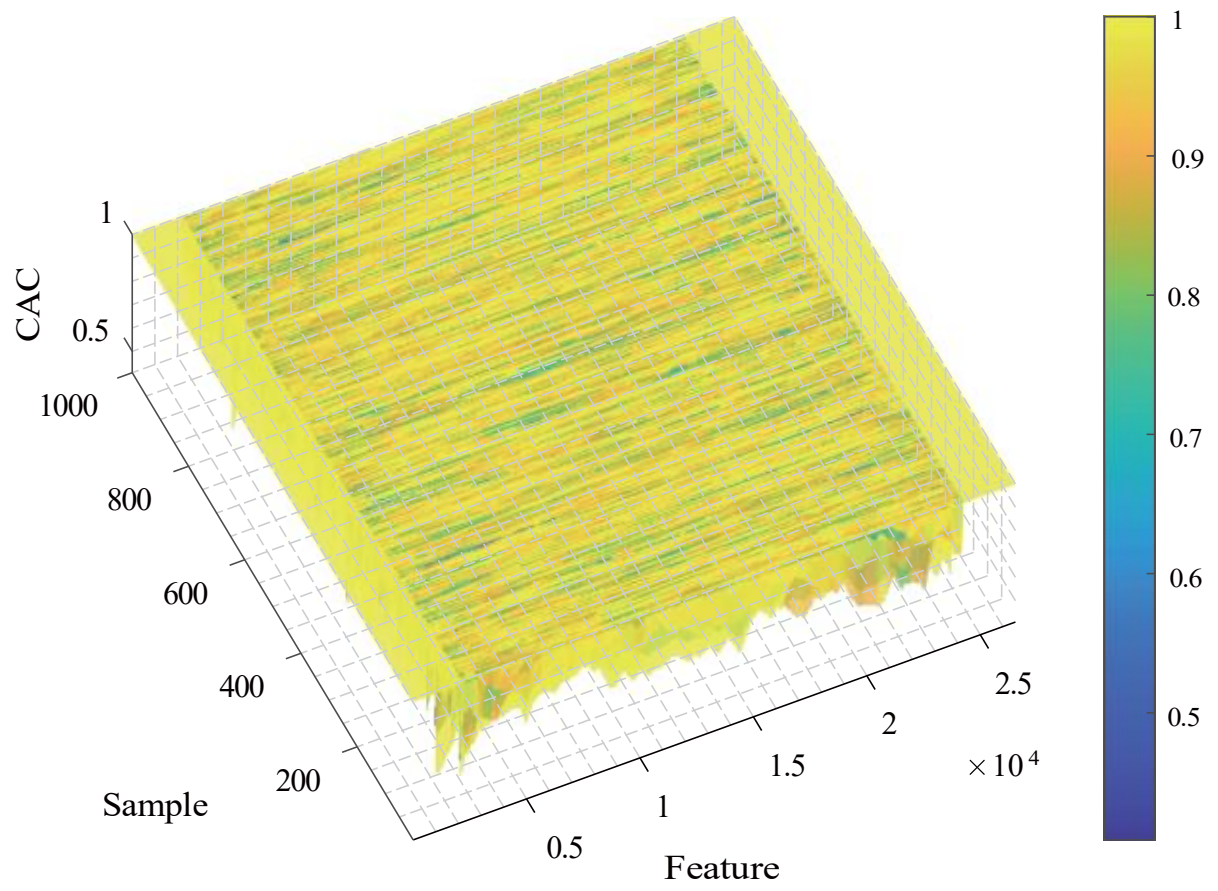

Figure 25: CAC profiles computed from 1,000 randomly generated observation sequences using data from the UNSW Bridge.

A second case is studied, in which the effectiveness of change point detection in identifying bridge state transitions is evaluated using an observation sequence that encompasses both benchmark and altered conditions. As described in Section 5, additional mass is added to the bridge to simulate a change in structural condition. An observation sequence, consisting of 20 samples from the benchmark condition and 10 from the damaged state, is established. This observation sequence is shown in the first row of Figure 26, with the corresponding CAC profile presented in the second row. A dashed line indicates the transition point between the two conditions. The CAC profile shows a clear minimum at the transition point, indicating that the method effectively detects the induced change.

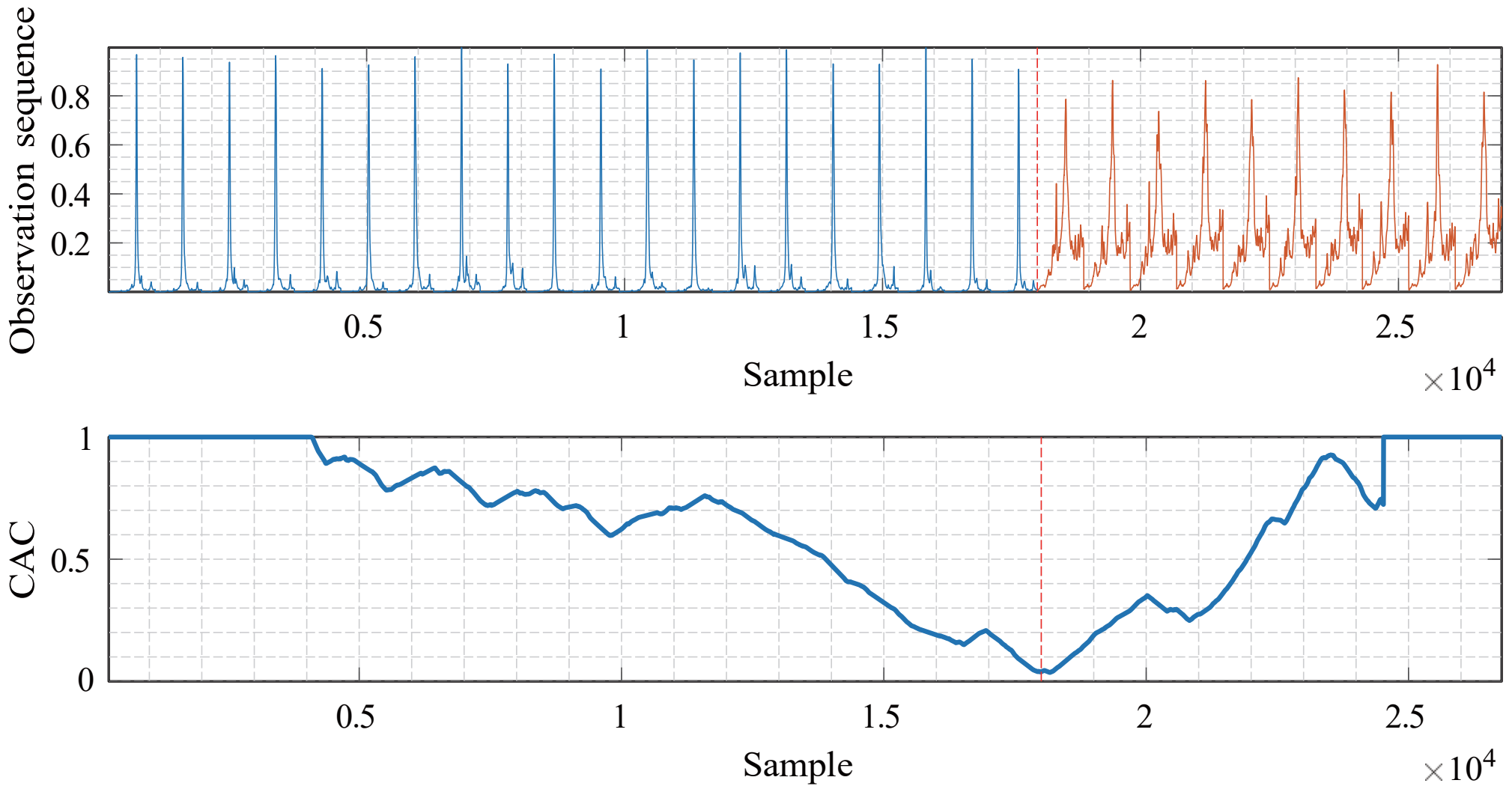

Figure 26: Observation sequence and corresponding CAC profile for the UNSW Bridge under benchmark and Changed conditions.

The consistency of the method in detecting change points is evaluated using 1,000 observation sequences generated from both benchmark and damaged conditions. The corresponding CAC values from all sequences are formed into the surface illustrated in Figure 27. This surface displays a clear line-shaped minimum aligned with the transition point across all sequences, indicating the location of change. This result demonstrates the reliability of the MP method in detecting bridge state changes across multiple inspections.

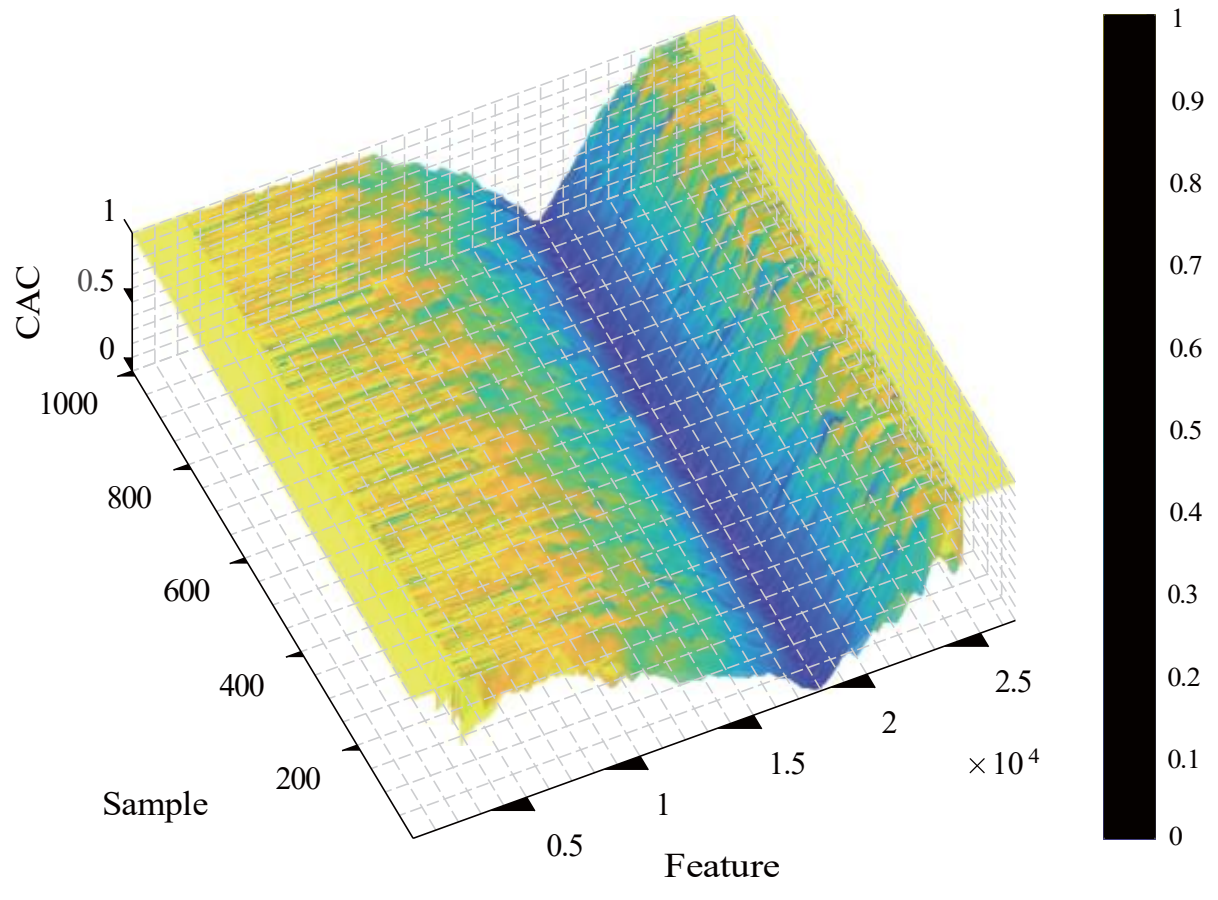

Figure 27: CAC profiles from 1,000 observation sequences, including both benchmark and damaged conditions for the UNSW Bridge.

### 6.2. Case Study II:

#### 6.2.1. Bridge Condition Assessment using AAE

Following the procedure described in Section 5.2, this section presents the application of the AAE-based methodology to the second case study. As previously mentioned, multiple vehicle crossings were conducted. From these measurements, the first singular values of the CPSD derived from the four acceleration responses, **S** was extracted and preprocessed, as detailed in Section 5.2. The resulting samples were then divided into training and testing sets using an 80:20 split, and the AAE model was trained accordingly. Figure 28 presents an example of the reconstruction of a **S** sample from the nominal state after training.

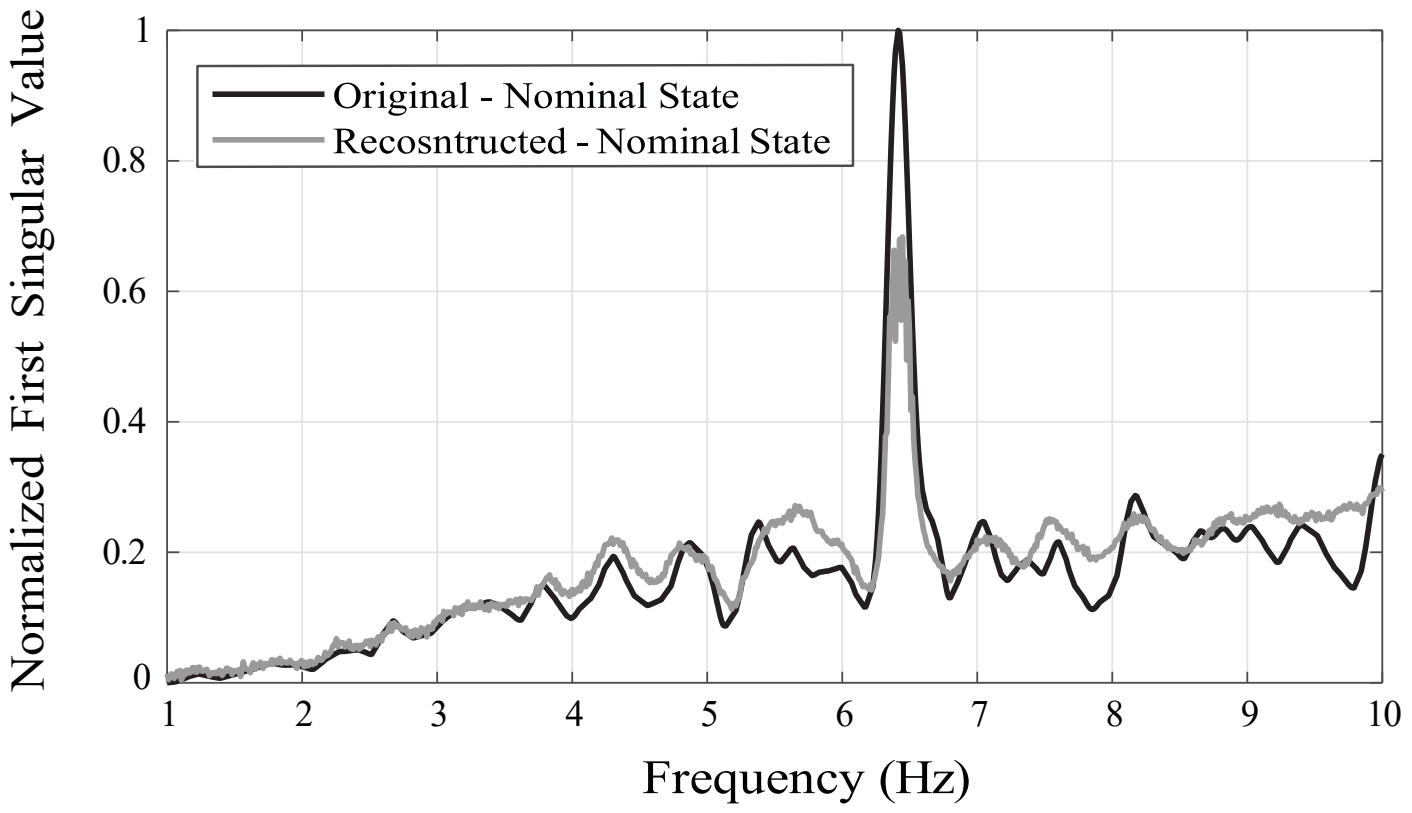

Figure 28: Reconstruction of a **S** sample using the AAE.

Once the AAE is trained and the damage threshold (see Section 5.2) is defined using the training set (i.e., 0.069), the test set is passed through the network. Given the high mass of the structure and its operational condition, it was challenging to add sufficient mass to alter its local dynamic behaviour. Therefore, this case study focuses solely on identifying the nominal bridge state, without considering modified or damaged states. Figure 29 shows the reconstruction errors corresponding to the nominal test set.

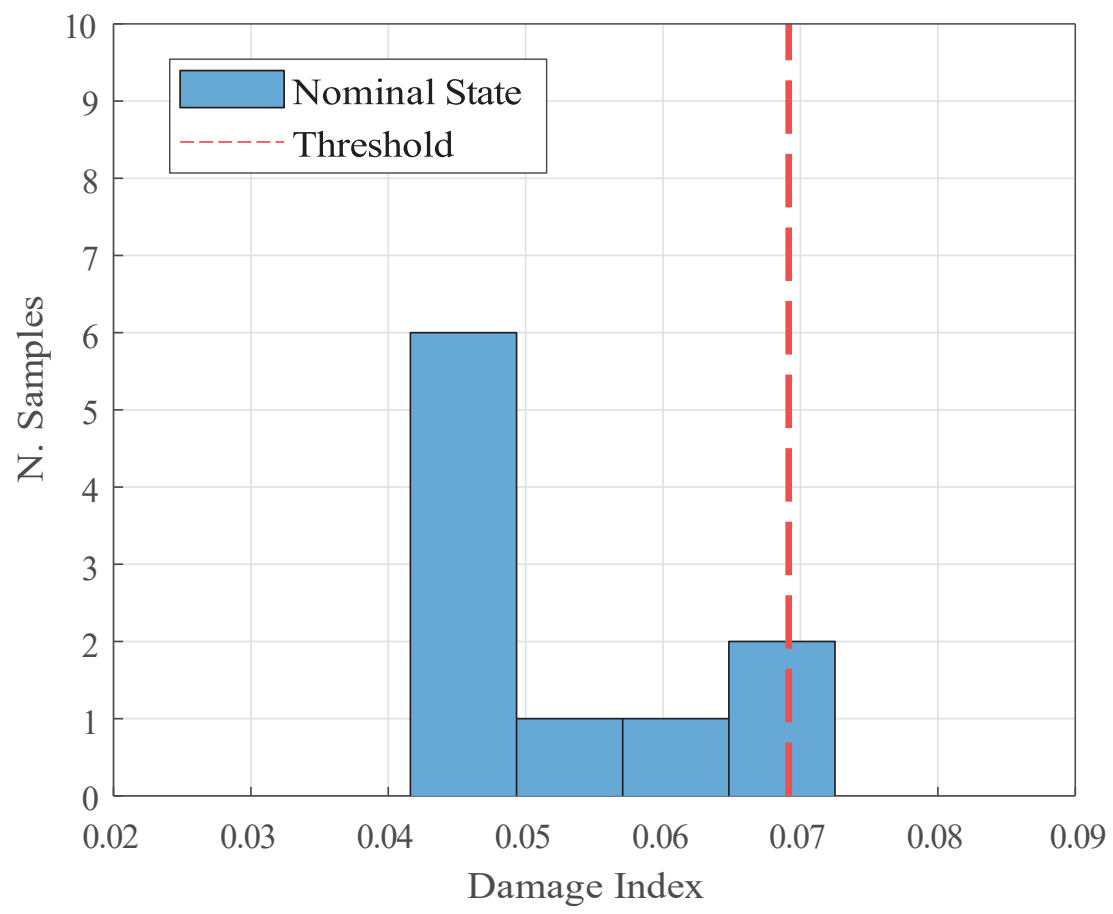

Figure 29: Reconstruction errors from the validation and test sets.

As illustrated in Figure 29, one out of ten samples in the test set of the nominal bridge condition exceeded the defined damage threshold. This result suggests that the AAE may not have fully learned the representative features of the benchmark state of the bridge. The limited number of available samples introduces uncertainty in the prediction and highlights the importance of using a larger dataset during the training phase. Although the method identified a false positive primarily due to the small number of samples used for training, the methodology still demonstrates potential for application in bridge anomaly detection by accurately reconstructing the healthy samples from the bridge.

#### *6.2.2. Bridge Condition Assessment using MP*

This section investigates the performance of the change point detection method presented in Section 5.3. The singular values obtained from the indirect approach presented in Section 4.2 are used to construct the observation sequences. To reduce the influence of noise, five singular value samples are randomly selected from a total of 39 samples and averaged to generate a new spectral sample. Each spectral sample contains 900 singular value points within the frequency range of 1 Hz to 10 Hz and is normalized between 0 and 1.

In this case study, the observation sub-sequence is constructed by assembling 30 samples, each generated by the averaging process mentioned previously. The complete subsequence, shown in the first row of Figure 30, represents a scenario in which all 30 inspections are conducted under benchmark condition.

The observation sequence has a length of 27,000 points (30 samples $\times$ 900 points). The MP method described in Section 5.3 is then applied to calculate the CAC. The resulting CAC is shown in the second row of Figure 30. Although minor deviations from the constant value of one are observed in the CAC curve, no significant decrease is present. Since the observation sequence does not contain any change in bridge condition, the CAC is expected to remain close to one. The result confirms this expectation, suggesting that the bridge condition remains consistent throughout the observation period.

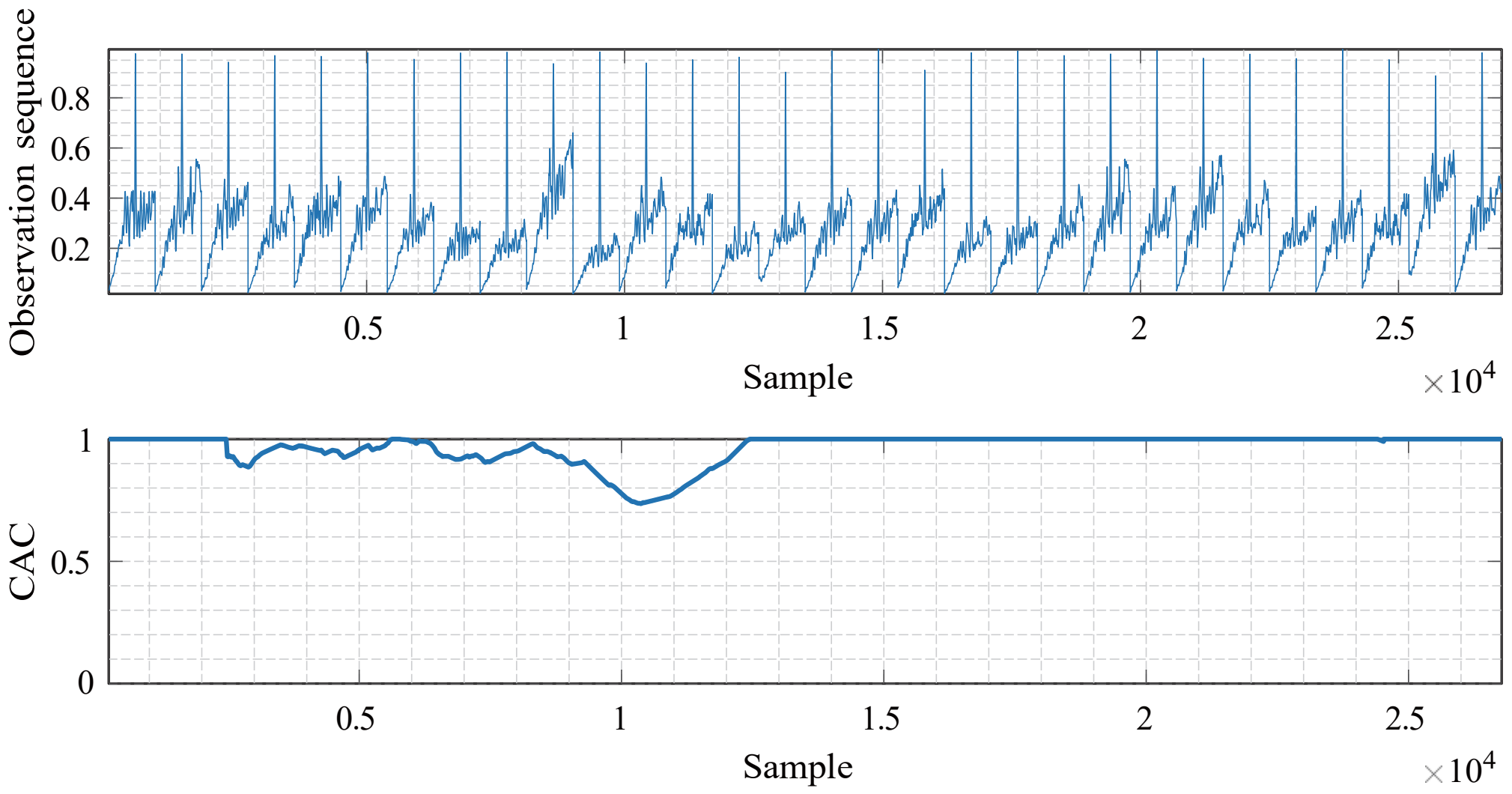

Figure 30: Observation sequence and corresponding CAC profile for the nominal condition of the Bulli Collie Bridge.

To further evaluate the robustness of the proposed method, 1,000 independent observation sequences are generated, and the corresponding CACs are computed. As shown in Figure 31, the CAC profiles form a surface with a value of one. Since the bridge condition remains unchanged in all observation sequences, no change points are expected, and the CAC reflects this by maintaining a maximum value across the surface.

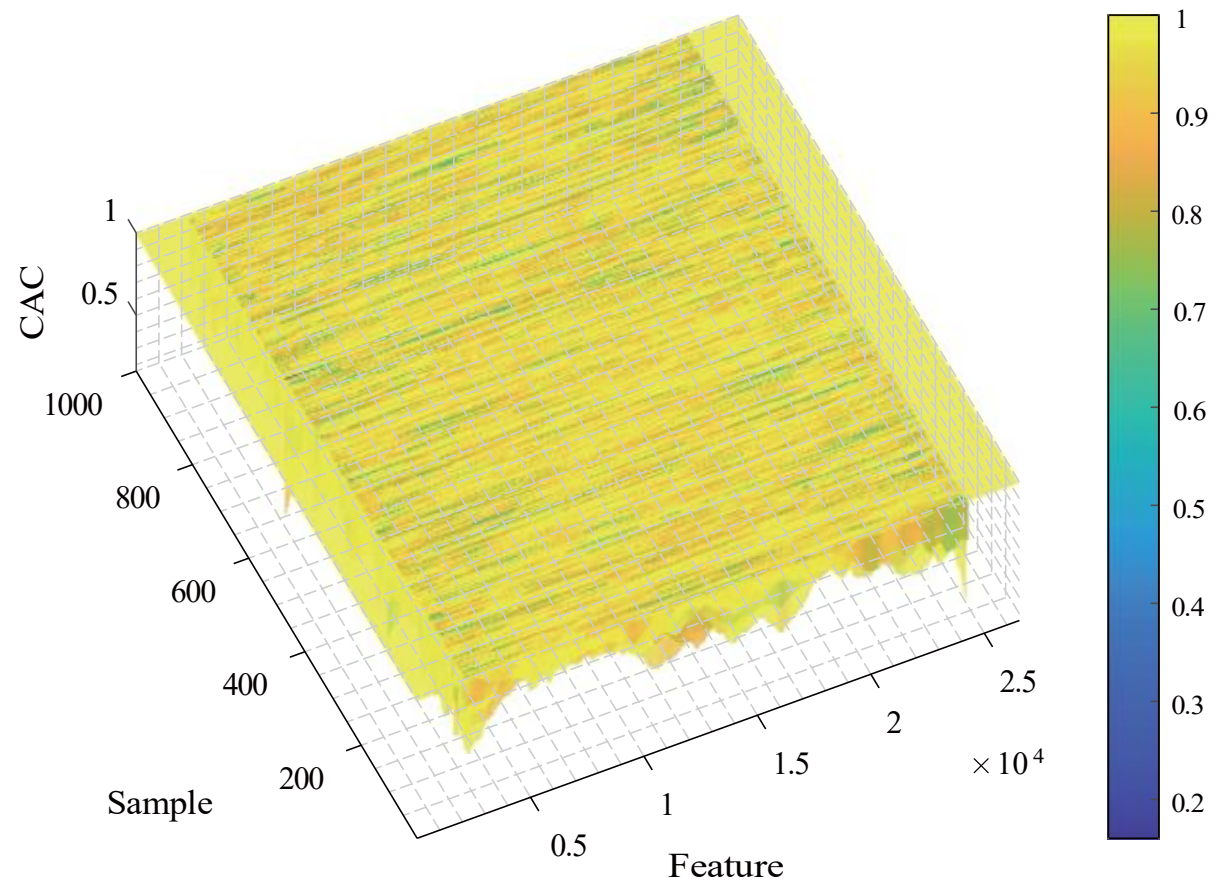

Figure 31: CAC profiles computed from 1,000 randomly generated observation sequences using data from the Bulli Colliery Bridge.

## 7. Conclusions

This study evaluated two previously developed drive-by inspection methods: the Adversarial Autoencoder and the Matrix Profile approaches. These methods were applied to data col-lected from two full-scale bridges using a custom-built electric inspection vehicle equipped with onboard accelerometers. The two case studies presented in this work correspond to the UNSW Bridge and the Bulli Colliery Bridge, respectively.

The first natural frequencies of both bridges were first estimated using both direct and indirect approaches. In both cases, Frequency Domain Decomposition was employed to extract the fundamental frequencies. Case Study I exhibited an average frequency of 6.65 Hz, while Case Study II showed an average of 6.42 Hz. These results confirm that the fundamental modal frequencies of the bridges can be accurately identified using drive-by bridge inspection techniques.

The AAE methodology was further investigated as a tool for assessing structural variation. In Case Study I, the nominal condition of the structure was intentionally altered by introducing localised added mass, enabling the simulation of damage and evaluation of the classification capability of the method. In contrast, Case Study II did not include damage simulation due to the high mass and operational constraints of the structure, which made it impractical to induce a measurable dynamic change. Instead, the focus was on characterising the nominal state and minimising false positives. The methodology successfully identified structural variation in Case Study I and accurately characterised the nominal state in Case Study II.

The MP approach was similarly applied to both case studies to detect change points within the observation sequences. When the bridges remained in their benchmark states, the corrected arc curve stayed close to one, with no significant minima observed, indicating that no false detections occurred.

In summary, this study demonstrates the feasibility of applying unsupervised learning and change point detection techniques for drive-by bridge inspection using a customisable inspection vehicle. The findings provide a solid foundation for future work in structural damage identification. One of the key advantages of the proposed inspection vehicle is its modular and customisable nature, which opens opportunities to explore the design space and optimise its configuration for enhanced sensing performance. Future research will also focus on studying long-term patterns and the effects of environmental and operational variability, as well as validating the methodology across a broader range of bridge types and conditions.

## 8. Acknowledgments

The authors would like to thank the Australian Research Council (ARC) for the provision of support under the Discovery Early Career Researcher Award (DECRA) scheme with grant number DE210101625. Additionally, this research was supported under the Australian Research Council's Industrial Transformation Research Programme (IH210100048). Thanks are also extended to Transport for NSW (TfNSW) for facilitating the field test on the Bulli Colliery Bridge.